\documentclass[a4paper,11pt]{article}
\pdfoutput=1
\usepackage{jheppub}

\newcommand{\ii}{\mathrm{i}}

\newcommand{\vev}[1]{\langle #1 \rangle}
\newcommand{\Tr}{\mathrm{Tr}\,}

\newcommand{\diag}{\mathrm{diag}\,}

\newcommand{\one}{{\rm 1\kern -.9mm l}}

\newcommand{\ft}[2]{{\textstyle\frac{#1}{#2}}}

\newcommand{\rme}{\mathrm{e}}

\newdimen\tableauside\tableauside=1.0ex
\newdimen\tableaurule\tableaurule=0.4pt
\newdimen\tableaustep
\def\phantomhrule#1{\hbox{\vbox to0pt{\hrule height\tableaurule
width#1\vss}}}
\def\phantomvrule#1{\vbox{\hbox to0pt{\vrule width\tableaurule
height#1\hss}}}
\def\sqr{\vbox{%
  \phantomhrule\tableaustep
\hbox{\phantomvrule\tableaustep\kern\tableaustep\phantomvrule\tableaustep}%
  \hbox{\vbox{\phantomhrule\tableauside}\kern-\tableaurule}}}
\def\squares#1{\hbox{\count0=#1\noindent\loop\sqr
  \advance\count0 by-1 \ifnum\count0>0\repeat}}
\def\tableau#1{\vcenter{\offinterlineskip
  \tableaustep=\tableauside\advance\tableaustep by-\tableaurule
  \kern\normallineskip\hbox
    {\kern\normallineskip\vbox
      {\gettableau#1 0 }%
     \kern\normallineskip\kern\tableaurule}%
  \kern\normallineskip\kern\tableaurule}}
\def\gettableau#1 {\ifnum#1=0\let\next=\null\else
  \squares{#1}\let\next=\gettableau\fi\next}
\def\XXint#1#2#3{{\setbox0=\hbox{$#1{#2#3}{\int}$}
     \vcenter{\hbox{$#2#3$}}\kern-.5\wd0}}

\title{\boldmath Chiral observables and S-duality in ${\mathcal N}=2^{\star}$ U$(N)$ gauge theories}

\author[a]{S.~K.~Ashok,}
\affiliation[a]{Institute of Mathematical Sciences \\
   C.~I.~T.~Campus, Taramani\\
   Chennai, India 600113}
\emailAdd{sashok@imsc.res.in}

\author[b]{M.~Bill\`o,}
\affiliation[b]{Universit\`a di Torino, Dipartimento di Fisica
\\ and I.~N.~F.~N.~- sezione di Torino, 
Via P. Giuria 1, I-10125 Torino, Italy}
\emailAdd{billo@to.infn.it}

\author[a]{E.~Dell'Aquila,}
\emailAdd{edellaquila@imsc.res.in}

\author[b]{M.~Frau,}
\emailAdd{frau@to.infn.it}

\author[c]{A.~Lerda,}
\affiliation[c]{Universit\`a del Piemonte Orientale, Dipartimento di Scienze e Innovazione Tecnologica, \\
and I.~N.~F.~N.~- sezione di Torino, Via P. Giuria 1, I-10125 Torino, Italy}
\emailAdd{lerda@to.infn.it}

\author[b]{M.~Moskovic,}
\emailAdd{moskovic@to.infn.it}

\author[a]{and M.~Raman}
\emailAdd{madhur@imsc.res.in}

\abstract{We study ${\mathcal N}=2^{\star}$ theories with gauge group U$(N)$ 
and use equivariant localization to calculate the quantum expectation values of the simplest chiral ring elements. 
These are expressed as an expansion in the mass 
of the adjoint hypermultiplet, with coefficients given by quasi-modular forms of 
the S-duality group. Under the action of this group, we construct combinations 
of chiral ring elements that transform as modular forms of definite weight. 
As an independent check, we confirm these results by comparing the spectral 
curves of the associated Hitchin system and the elliptic Calogero-Moser system. 
We also propose an exact and compact expression for the 1-instanton contribution 
to the expectation value of the chiral ring elements.}


\begin{document} 
\maketitle
\flushbottom

\section{Introduction}
\label{sec:intro}
$\mathcal{N} = 2$ super Yang-Mills (SYM) theories in four dimensions are 
an extraordinarily fertile ground to search for exact results.
Indeed, their non-perturbative behaviour can be tackled both via the Seiberg-Witten (SW)
description of their low-energy effective theory \cite{Seiberg:1994rs,Seiberg:1994aj},
and via the microscopic computation of instanton effects by means of localization techniques 
\cite{Nekrasov:2002qd}--\nocite{Flume:2002az,Nekrasov:2003rj,Bruzzo:2002xf, Losev:2003py}\cite{Flume:2004rp}. 
Understanding the far-reaching consequences of strong/weak coupling 
dualities in the effective theory has always been a crucial ingredient in the SW approach. 
On the other hand, the same dualities can also be exploited 
in the microscopic description through the associated modular structure.
The comparison of how these dualities may be used to constrain physical observables in the two approaches is one of the main themes of this paper.

Among the $\mathcal{N} = 2$ models, much effort has been devoted to gaining a deeper understanding of 
superconformal theories and their massive deformations (see for example the collection of reviews 
\cite{Teschner:2014oja} and references therein), where many different approaches have been investigated. Among these we can mention the relation to integrable models \cite{Nekrasov:2009rc}, 
the 2d/4d AGT correspondence \cite{Alday:2009aq,Alday:2009fs},
the use of matrix model techniques \cite{Dijkgraaf:2009pc,Cheng:2010yw} and the link
to topological string amplitudes through geometric engineering \cite{Antoniadis:2010iq}--\nocite{Huang:2011qx}\cite{Florakis:2015ied}. Furthermore, the pioneering work of Gaiotto \cite{Gaiotto:2009we} has taught us that the duality properties are of the utmost relevance.

In this paper we focus on $\mathcal{N}=2^\star$ theories, which we briefly review in Section~\ref{secn:rev}. Besides the gauge vector multiplet, they contain an adjoint hypermultiplet of mass $m$ that interpolates 
between the $\mathcal{N} = 4$ SYM theories (when $m\to 0$)
and the pure $\mathcal{N} = 2$ SYM theories (when $m\to \infty$). 
The $\mathcal{N}=2^\star$ theories inherit from the $\mathcal{N} = 4$ models an interesting action 
of the S-duality group; 
in particular, their prepotential satisfies a modular anomaly equation which greatly constrains its form.
Modular anomaly relations in gauge theories were first noticed in \cite{Minahan:1997if} and are related to 
the holomorphic anomaly equations that occur in topological string theories on local Calabi-Yau manifolds \cite{Bershadsky:1993ta}\nocite{Witten:1993ed,Aganagic:2006wq}--\cite{Gunaydin:2006bz}. These equations have been studied in a variety of settings, for example in an $\Omega$ background
\cite{Huang:2006si}--\nocite{Grimm:2007tm,Huang:2009md,Huang:2010kf,Huang:2011qx,Galakhov:2012gw,Billo:2013fi,Billo:2013jba,Nemkov:2013qma,Billo:2014bja, Lambert:2014fma, Beccaria:2016wop}\cite{Beccaria:2016nnb},  from the point of view of the AGT correspondence
\cite{Marshakov:2009kj}--\nocite{KashaniPoor:2012wb,Kashani-Poor:2013oza}\cite{Kashani-Poor:2014mua}, 
in the large-$N$ limit \cite{Billo:2014bja}, and in SQCD models with fundamental matter \cite{Billo:2013fi,Billo:2013jba,Ashok:2015cba,Ashok:2016oyh}.

Recently, the modular anomaly equation for $\mathcal{N}=2^\star$ theories with arbitrary gauge groups has been linked in a direct way to S-duality \cite{Billo':2015ria}--\nocite{Billo':2015jta}\cite{Billo:2016zbf}. 
This approach has led to a very efficient way of determining the mass expansion of the prepotential in terms of:  $i)$ quasi-modular functions of the gauge coupling and $ii)$ the vacuum expectation values $a_u$ of the scalar field $\Phi$ of the gauge multiplet such that only particular combinations, defined purely in terms of sums over the root lattice of the corresponding Lie algebra, appear. These results have been checked against explicit computations using equivariant localization. 

In this work, we take the first steps towards showing that similar modular structures also exist 
for other observables of ${\mathcal N} =2^\star$ gauge theories. 
We choose to work with U$(N)$ gauge groups, and consider the quantum expectation values 
\begin{equation}
\label{Ovev}
\vev{\Tr \Phi^n}~.
\end{equation}
The supersymmetry algebra implies that correlators of chiral operators factorize and can therefore be expressed in terms of the expectation values in \eqref{Ovev}.\footnote{More general correlators involving also one anti-chiral operator have recently been considered in \cite{Gerchkovitz:2016gxx}.}

A priori, it is not obvious that these chiral observables exhibit modular behaviour. However, we show 
that it is always possible to find combinations that transform as modular forms of definite weight under 
the non-perturbative duality group SL$(2,\mathbb{Z})$. These combinations have a natural interpretation as modular-covariant coordinates on the Coulomb moduli space, and can be analysed using two different techniques: $i$) the SW approach via curves and differentials, 
and $ii$) equivariant localization combined with the constraints arising from S-duality.

For ${\mathcal N}=2^{\star}$ theories there are many distinct forms of the SW curve 
that capture different properties of the chiral observables. In one approach, due to Donagi and Witten \cite{Donagi:1995cf,Donagi:1997sr}, the SW curve has coefficients $A_n$
that have a natural interpretation as modular-covariant coordinates on the Coulomb moduli space. Thus, this approach provides us with a natural setting to study the elliptic and modular properties of the observables (\ref{Ovev}). Another form of the SW curve was found by using the relation with integrable systems \cite{Gorsky:1995zq}. For the ${\mathcal N}=2^{\star}$ theory, the relevant curve was proposed by D'Hoker and Phong \cite{D'Hoker:1997ha,D'Hoker:1999ft}, who used the close relation between the gauge theory and the elliptic Calogero-Moser system \cite{Martinec:1995qn}. 
In this second formulation, the coefficients of the spectral curve of the integrable system are interpreted as symmetric polynomials built out of the quantum chiral ring elements \eqref{Ovev}. A third form of the
SW curve for the ${\mathcal N}=2^{\star}$ theories was proposed by Nekrasov and Pestun \cite{Nekrasov:2012xe} together with an extension to general quiver models.
In Section \ref{secn:curves} we review and relate the first two descriptions of the SW curve which are suitable for our purposes. 
This comparison will lead to interesting relationships between the coefficients of the respective curves. Along the way, we will find it necessary to modify the analysis of \cite{Donagi:1995cf} in a subtle but important way.

It is clearly desirable to work with chiral observables that in the classical limit coincide with the symmetric polynomials built out of the vacuum expectation values $a_u$. As we discuss 
in Section~\ref{secn:periodsandmodularity},
this can be done in two ways. 
The first is to compute the period integrals in the Donagi-Witten form of the curve
as a series expansion in the mass $m$ of the adjoint hypermultiplet. Inverting this expansion 
order by order in $m$ gives us an expression for the $A_n$ in terms of the $a_u$. The second way is to postulate that the $A_n$ have a definite modular weight under the S-duality group, and use the well-understood action of S-duality to derive a modular anomaly equation that recursively determines them up to modular pieces. In this derivation, it is crucial that the prepotential and hence the dual periods of the ${\mathcal N}=2^{\star}$ theory are known in terms of quasi-modular forms.
In both ways it turns out that the chiral observables can be expressed in terms of quasi-modular forms and of particular functions of the $a_u$ involving only sums over the weight and root lattices of the Lie 
algebra $\mathfrak{u}(N)$, generalizing those appearing in the prepotential. 
%{From} these explicit expressions one can check that indeed the $A_n$'s have weight $n$ under the modular group.

In Section \ref{secn:localization} we test our findings against explicit microscopic computations of the observables (\ref{Ovev}) using equivariant localization techniques 
\cite{Nekrasov:2002qd}--\nocite{Flume:2002az,Nekrasov:2003rj,Bruzzo:2002xf, Losev:2003py}\cite{Flume:2004rp} (for further technical details see also \cite{Billo:2012st}).
We find that the chiral observables computed using localization can be matched with those obtained from the SW curves by a redefinition of the chiral ring elements. Such a redefinition contains only a finite number of terms and is exact both in the mass of the hypermultiplet and in the gauge coupling. It is well known that the localization results for the chiral observables do not, in general, satisfy the classical chiral ring relations \cite{Cachazo:2002ry, Flume:2004rp, Nekrasov:2012xe}.   
Strikingly, we show that the redefinition of the chiral ring elements which allow the matching of the two sets of results can be interpreted as a judicious choice of coordinates on the Coulomb moduli space in which the {\it classical} chiral ring relations are naturally satisfied. 

In Section \ref{secn:one-inst}, we focus on the $1$-instanton contributions and, just as it was done 
for the prepotential in \cite{Billo':2015ria,Billo':2015jta}, we manage to
resum the mass expansion to obtain an exact expression involving only sums over roots and weights of the corresponding Lie algebra. 

Finally, we present our conclusions in Section~\ref{concl} and collect
various technical details in the appendices.

\section{\boldmath Brief review of ${\mathcal N}=2^{\star}$ U($N$) SYM theories}
\label{secn:rev}
The ${\mathcal N}=2^{\star}$ SYM theories are massive deformations of 
the ${\mathcal N}=4$ SYM theories arising when the adjoint hypermultiplet 
is given a mass $m$. The classical vacua of these theories on the Coulomb branch
are parametrized by the expectation values of the scalar field $\Phi$ in 
the vector multiplet, which in the U$(N)$ case is 
\begin{equation}
\vev{\Phi} \,\equiv\, a= \diag(a_1, a_2, \ldots, a_N) ~.
\label{au}
\end{equation}
When the complex numbers $a_u$ are all different, the gauge group is 
broken to its maximal torus U$(1)^N$. 
The low-energy effective action of this abelian theory is completely 
determined by a single 
holomorphic function $F(a)$, called the prepotential.
It consists of a classical term
\begin{equation}
F_{\text{class}} = \ii \pi \tau\,a^2\,\equiv\,\ii \pi \tau\sum_{u=1}^N a_u^2 ~,
\label{Fcl}
\end{equation}
where $\tau$ is the complexified gauge coupling 
\begin{equation}
\tau = \frac{\theta}{2\pi} + \ii \,\frac{4\pi}{g^2} ~,
\label{tau}
\end{equation}
and a quantum part 
\begin{equation}
f=F_{\text{1-loop}} + F_{\text{inst}}
\label{Fq}
\end{equation}
accounting for the $1$-loop and instanton corrections.

The $1$-loop term $F_{1\text{-loop}}$ is $\tau$-independent and takes 
the simple form (see for instance \cite{D'Hoker:1999ft})
\begin{equation}
F_{\text{1-loop}} 
= -\frac{1}{4}\sum_{\alpha\in\Psi}\left[(\alpha \cdot a)^2\log\left(\frac{\alpha\cdot a}{\Lambda}\right)^2 
-(\alpha\cdot a+m)^2 \log\left(\frac{\alpha\cdot a+m}{\Lambda}\right)^2\right]~,
\label{F1loop}
\end{equation}
where $\Lambda$ is an arbitrary scale and $\alpha$ is an element of the root 
system $\Psi$ of the gauge algebra. The first and seconds terms in (\ref{F1loop}) 
are, respectively, contributions from the vector multiplet and the massive hypermultiplet.

The instanton corrections to the prepotential are proportional to $q^k$, where
\begin{equation}
q=\rme^{2\pi\ii\tau}
\label{q}
\end{equation}
is the instanton counting parameter and $k$ is the instanton number.
These non-perturbative terms can be calculated either using the SW curve 
and corresponding holomorphic differential $\lambda_{\text{SW}}$ 
\cite{Seiberg:1994rs,Seiberg:1994aj}, or by a microscopic evaluation 
of the prepotential using localization 
\cite{Nekrasov:2002qd}--\nocite{Flume:2002az,Nekrasov:2003rj,Bruzzo:2002xf, Losev:2003py}\cite{Flume:2004rp}.

In the SW approach, besides the ``electric" variables $a_u$, one 
introduces dual or ``magnetic" variables defined by
\begin{equation}
a^{\text{D}}_u = \frac{1}{2\pi \ii}\, \frac{\partial F}{\partial a_u}~.
\label{adual}
\end{equation}
The pairs $(a_u, a_u^{\text{D}})$ describe the period 
integrals of the holomorphic differential $\lambda_{\text{SW}}$ over cycles of the Riemann 
surface defined by the SW curve. More precisely, one has
\begin{equation}
a_u = \oint_{A_u} \lambda_{\text{SW}} \qquad \text{and} \qquad a^{\text{D}}_u = \oint_{B_u} \lambda_{\text{SW}}~.
\label{vevperiods}
\end{equation}
Here, the $A$- and $B$-cycles form a canonically conjugate symplectic basis of cycles 
with intersection matrix $A_u \cap B_v = \delta_{uv}$.

For the ${\mathcal N}=2^{\star}$ U($N$) theory, the non-perturbative 
S-duality group has a simple embedding into the symplectic duality 
group Sp$(4N, \mathbb{Z})$ of the Riemann surface. In particular, 
the $S$-transformation acts by exchanging electric and magnetic variables, 
while inverting the coupling constant, namely
\begin{equation}
S(a_u) = a^{\text{D}}_u \, , \qquad S(a^{\text{D}}_v) = -a_v \, , \qquad S(\tau) = -\frac{1}{\tau} ~.
\label{Sona}
\end{equation}
Along with the $T$-transformation, given by
\begin{equation}
T(a_u) = a_u \, , \qquad T(a^{\text{D}}_v) = a^{\text{D}}_v + a_v\, ,\qquad T(\tau) = \tau + 1 ~,
\end{equation}
one generates the modular group SL$(2,\mathbb{Z})$. 

To discuss the $\mathcal{N}=2^\star$ prepotential and the action of 
the duality group on it, it is convenient to organize its quantum 
part (\ref{Fq}) as an expansion in powers of the hypermultiplet mass, as
\begin{equation}
f= \sum_{n=1}^{\infty} f_{n}\,m^{2n}~.
\label{fn}
\end{equation}
Notice that only even powers of $m$ occur in this expansion as a 
consequence of the $\mathbb{Z}_2$ symmetry that sends $m\to -m$. 
In order to write the coefficients $f_n$ in a compact form, it is 
useful to introduce the following lattice sums
\begin{equation}
\begin{aligned}
C^{\,p}_{\,n;m_1\cdots\, m_\ell}&=\sum_{\lambda\in\mathcal{W}}~\sum_{\alpha\in\Psi_\lambda}\,\sum_{\beta_1\not=\cdots\not=\beta_\ell\in\Psi_{\alpha}}
\frac{(\lambda\cdot a)^p}{(\alpha\cdot a)^n(\beta_1\cdot a)^{m_1}\cdots(\beta_\ell\cdot a)^{m_\ell}}
\end{aligned}
\label{Cpnm}
\end{equation}
where $\mathcal{W}$ is the set of weights $\lambda$ of the 
fundamental representation of 
U($N$), while $\Psi_\lambda$ and $\Psi_{\alpha}$ are the subsets 
of the root system $\Psi$ defined, respectively, by
\begin{equation}
{\Psi}_\lambda=\big\{\alpha\in {\Psi} \,\big|\,\lambda\cdot \alpha=1\big\}~,
\label{psilambda}
\end{equation}
for any $\lambda\in \mathcal{W}$, and by
\begin{equation}
\Psi_{\alpha}=
\big\{\beta\in \Psi \,\big|\,\alpha\cdot\beta=1\big\}~,
\label{psilambdaalpha}
\end{equation}
for any $\alpha\in\Psi$. Notice that 
\begin{equation}
C^{\,0}_{\,n;m_1\cdots \,m_\ell} = C_{n;m_1\cdots\, m_\ell}
\label{C0nm}
\end{equation}
where $C_{n;m_1\cdots m_\ell}$ are the lattice sums introduced 
in \cite{Billo':2015ria,Billo':2015jta,Billo:2016zbf}. Furthermore, we have
\begin{equation}
C^{\,\ell}_{\,0;0\cdots\, 0} = \sum_{u=1}^N a_u^{\,\ell}\,\equiv\, C^{\,\ell}~.
\label{Cell}
\end{equation}
Using this notation, the first few coefficients in the mass expansion of the 
U($N$) prepotential were shown to be given by \cite{Billo':2015ria}\,%
\footnote{We warn the reader that, for later convenience, we have changed notation with respect to 
\cite{Billo':2015ria} and have explicitly factored out the mass-dependence. So,
$f_n^{\text{there}}= f_n^{\text{here}}\,m^{2n}$.}
\begin{equation}
\begin{aligned}
f_1&=\frac{1}{4} \sum_{\alpha\in\Psi}  \log\left(\frac{\alpha\cdot a}{\Lambda}\right)^2~,\\
f_2&= -\frac{1}{24} \,E_2 \, C^{\,0}_{\,2} ~,\\
f_3&= -\frac{1}{720}\,\big(5 E_2^2 + E_4\big)\, C^{\,0}_{\,4} -
\frac{m^6}{576}\,\big(E_2^2 - E_4\big)\, C^{\,0}_{\,2;11}~,
\end{aligned}
\label{fnexpl}
\end{equation}
where $E_{2k}$ are the Eisenstein series (see Appendix \ref{appeisen}). These formulas encode
the \emph{exact} dependence on the coupling constant $\tau$. Indeed, by expanding the Eisenstein 
series in powers of $q$, one can recover the perturbative contributions, corresponding to the terms 
proportional to $q^0$, and the $k$-instanton contributions proportional to $q^k$.
Analogous expressions can be obtained for the higher order mass terms in the U($N$) theory and for 
other gauge algebras as well \cite{Billo':2015ria,Billo':2015jta,Billo:2016zbf}.

As discussed in great detail in \cite{Billo:2013fi, Billo:2013jba, Billo:2014bja} 
the prepotential coefficients $f_n$ satisfy the recursion relation
\begin{equation}
\label{rec}
\frac{\partial f_n}{\partial E_2} = -\frac{1}{24} \sum_{m=1}^{n-1}
\frac{\partial f_m}{\partial a}\cdot \frac{\partial f_{n-m}}{\partial a}~,
\end{equation}
which in turn implies that the quantum prepotential $f$ obeys the non-linear differential equation
\begin{equation}
\label{recdiff}
\frac{\partial f}{\partial E_2} +\frac{1}{24} \left(\frac{\partial f}{\partial a}\right)^2=0~.
\end{equation}
This equation, which is a direct consequence of the S-duality action (\ref{Sona}) on 
the prepotential, is referred to as the modular anomaly equation since $E_2$ has an anomalous modular behavior
\begin{equation}
E_2\left(-\frac{1}{\tau}\right)=\tau^2\Big(E_2(\tau)+\frac{6}{\ii\pi\tau}\Big)~.
\label{E2s}
\end{equation}

\section{\boldmath Seiberg-Witten curves for the ${\mathcal N}=2^{\star}$ U($N$) SYM theories}
\label{secn:curves}
In this section we review and compare two distinct algebraic approaches to describe the low-energy effective
quantum dynamics of the ${\mathcal N}=2^{\star}$ U$(N)$ SYM theory.
The first approach is due to Donagi and Witten \cite{Donagi:1995cf} (see also \cite{Donagi:1997sr}), 
while the second approach is due to D'Hoker and Phong \cite{D'Hoker:1997ha}. 
Even though some of the following considerations already appeared in the literature \cite{Itoyama:1995nv, Ennes:1999fb},
we are going to revisit the comparison between the two curves with the purpose of introducing 
the essential ingredients for the non-perturbative analysis presented in later sections.

\subsection{The Donagi-Witten curve}
\label{DW}

In this first approach, the algebraic curve of the ${\mathcal N}=2^{\star}$ U$(N)$ theory is 
given as an $N$-fold cover of an elliptic genus-one curve. The latter takes the standard Weierstra\ss~form
\begin{equation}
y^2 = (x-e_1)(x-e_2)(x-e_3)~,
\label{curve1}
\end{equation}
where the $e_i$ sum to zero and their differences are given in terms of the Jacobi $\theta$-constants 
\cite{Seiberg:1994aj} as\,%
\footnote{We use a different notation and normalization as compared to 
Ref.~\cite{Seiberg:1994aj}. In particular our normalizations are such that the 
$\alpha$-period of the uniformizing coordinate of the torus is $\omega_1=2\pi\ii$. }
\begin{equation}
e_2-e_3 = \frac{1}{4}\,\theta_2(\tau)^4~, \qquad e_2-e_1 = \frac{1}{4}\,\theta_3(\tau)^4 ~,\qquad
e_3-e_1 = \frac{1}{4}\,\theta_4(\tau)^4~.
\label{ei}
\end{equation}
Here $\tau$ is the complex structure parameter of the elliptic curve which is identified with the gauge coupling
(\ref{tau}) and the $\theta$-constants have the following Fourier expansions
\begin{equation}
\theta_2(\tau) = \sum_{n\in \mathbb{Z}} q^{\frac{1}{2}\left(n-\frac{1}{2}\right)^2}~,\qquad
\theta_3(\tau) = \sum_{n\in \mathbb{Z}} q^{\frac{1}{2}n^2}~,\qquad
\theta_4(\tau) = \sum_{n\in \mathbb{Z}} (-1)^n\, q^{\frac{1}{2}n^2}~,
\label{thetas}
\end{equation}
where $q$ is as in (\ref{q}). Using the relations between the $\theta$-constants and the Eisenstein series (see
\ref{e4e6theta}), the elliptic curve (\ref{curve1}) can be rewritten as
\begin{equation}
\label{ellipticW}
y^2 = x^3-\frac{E_4}{48}\,x  +\frac{E_6}{864}~.
\end{equation}
Since $E_4$ and $E_6$ are
modular forms of weight $4$ and $6$, for consistency $x$ and $y$ must have modular 
weight $2$ and $3$ respectively. If we recall the uniformizing solution
in terms of the Weierstra\ss~function $\wp(z)$, which obeys
\begin{equation}
\wp'(z)^2 = 4 \,\wp^3(z) -\frac{4\pi^4\,E_4}{3}\, \wp(z) -\frac{8\pi^6\,E_6}{27}
\label{PW}
\end{equation}
when $z\sim z+1$ and $z\sim z+\tau$,
then by comparing with (\ref{ellipticW}) we straightforwardly obtain the following identifications:
\begin{equation}
2\,y= \frac{\wp'(z)}{(2\pi\ii)^3}~,\quad\qquad x = \frac{\wp(z)}{(2\pi\ii)^2} ~.
\label{xyvswp}
\end{equation}

In this framework, the curve of the ${\mathcal N}=2^{\star}$ U($N$) theory is described by
the equation 
\begin{equation}
F(t,x,y) = 0
\label{DWcurve0}
\end{equation}
where $F(t,x,y)$ is a polynomial of degree $N$. Modular covariance is extended 
to this equation by assigning modular weight $1$ to the variable $t$. 
Certain technical conditions described in detail in \cite{Donagi:1995cf, Donagi:1997sr} 
allow one to fix the form of $F$ to be
\begin{equation}
\label{DWcurve}
F(t,x,y) = \sum_{n=0}^N \, (-1)^n\, A_n\, P_{N-n}(t,x,y)\,,
\end{equation}
where $A_0=1$ and the remaining $N$ quantities $A_n$ parametrize the Coulomb branch of 
the moduli space. The polynomials $P_n(t,x,y)$ are of degree $n$ and are \emph{almost} 
completely determined by the recursion relations \cite{Donagi:1995cf}
\begin{equation}
\frac{\text{d}P_n}{\text{d}t} = n\, P_{n-1} \, ,
\label{Pn}
\end{equation}
combined with physical requirements related to the behaviour of $F$ in the limits $x,y \rightarrow \infty$. 

At the first two levels, $n=0$ and $n=1$, in view of the weights assigned to $x$ and $y$, 
the polynomials are uniquely fixed to be
\begin{equation}
P_0 =1~,\qquad
P_1 =t~.
\label{P01}
\end{equation}
At the next order, $n=2$, the solution to the recursion equation (\ref{Pn}) is
\begin{equation}
P_2 = t^2 +c\, m^2
\end{equation}
where the second term is an integration constant depending on the hypermultiplet mass
that is allowed since $P_2$ has mass dimension $2$. In addition, since $P_2$ has 
modular weight $2$, the coefficient $c$ must be an elliptic or modular function of 
weight $2$. There is a unique such function, namely $x$, and thus $P_2$ must be of the form
\begin{equation}
P_2 = t^2 + \alpha\, x\,m^2
\label{P2}
\end{equation}
where $\alpha$ is a numerical coefficient which is fixed by requiring a specific behavior at
infinity \cite{Donagi:1995cf}. 

If we choose coordinates such that $u=0$ parametrizes the point at infinity, then taking into account
that $x$ is an elliptic function of weight $2$, we can write
\begin{equation}
x = \frac{1}{u^2}~.
\label{xu}
\end{equation}
In terms of this variable, the required behavior at infinity is that under the shift
\begin{equation}
t~\to~t+\frac{m}{u}~,
\end{equation}
the function $F$, and therefore all polynomials $P_n$, must have at most a \emph{simple} pole in $u$,   
namely for $u\to 0$ they must behave as\,%
\footnote{This follows from the 
requirement in \cite{Donagi:1995cf} that the adjoint scalar field $\Phi$ 
has the following behaviour near the point $u=0$ on the torus:
\begin{equation*}
\Phi = \frac{m}{u}\, \text{diag}(1,1, \ldots, -(N-1))+ \text{regular terms} \,.
\end{equation*}
The residue $m$ is identified with the mass of the adjoint hypermultiplet. The function 
$F(t,x,y)$, which defines the $N$-fold spectral cover of the torus, 
is identified with the equation $\det (t\mathbf{1}-\Phi)=0$. 
The shift in $t$ above ensures that $N-1$ of the eigenvalues of $\Phi$ have 
no pole as $u\rightarrow 0$ and this is what constrains the growth 
of the polynomials $P_n$ near infinity (see \cite{Donagi:1995cf} for more details).}
\begin{equation}
P_n\Big(t+\frac{m}{u}\Big) \sim \frac{\alpha_n}{u} +\mathrm{regular}~.
\end{equation}
The requirement that all higher order poles in $u$ cancel constrains the integration constants that 
are allowed to appear. For example, imposing this behavior, one can easily fix the constant 
$\alpha$ in (\ref{P2}) and find that final form of $P_2$ is
\begin{equation}
P_2 = t^2 -\, x\,m^2~.
\label{P2new}
\end{equation}
To fix the higher order polynomials, it is necessary to know the 
behaviour of $y$ near $u=0$. Using the algebraic equation (\ref{ellipticW}), we easily find
\begin{equation}
y = \frac{1}{u^3} \sqrt{1-\frac{E_4}{48}u^4+\frac{E_6}{84}u^6} \,
= \,\frac{1}{u^3}-\frac{E_4}{96}\,u - \frac{E_6}{1728}\,u^3+ \cdots~.
\end{equation}
Using this and (\ref{xu}), we can completely determine the polynomial $P_3$ and get
\begin{equation}
P_3=t^3 -3\,t\,x\, m^2+2\,y\,m^3~.
\end{equation}
However, at the next level, we find that 
\begin{equation}
\label{P4new}
P_4 =t^4-6\,t^2\,x\,m^2+8\,t\,y\,m^3-\left(3\,x^2-\alpha\,E_4\right)m^4
\end{equation}
satisfies all requirements for \emph{any} value of $\alpha$. In \cite{Donagi:1995cf, Donagi:1997sr} the simplest
choice $\alpha=0$ was made, but we will find that it is actually essential to keep the $\alpha$-dependence
and fix it to a different value.

This procedure can be iterated without any difficulty and in Appendix \ref{DWpolygeneral} 
we list a few of the higher degree polynomials $P_n$ that we find in this way. They 
differ from the ones listed in \cite{Donagi:1995cf, Donagi:1997sr} 
by elliptic and modular functions. At first glance, these might seem trivial 
modifications since, for example in \eqref{P4new}, the difference is 
proportional to $E_4$, which is a modular form of weight $4$. However, for $\alpha\ne 0$, 
this new term feeds into the iterative procedure to calculate the higher $P_n$, which 
in turn depend on these coefficients. These modified higher degree polynomials will 
play a crucial role in the following. 

Using the explicit form of the polynomials $P_n$ given in Appendix~\ref{DWpolygeneral} 
and collecting the powers of $t$, we find that the curve equation (\ref{DW}) is
\begin{equation}
\begin{aligned}
F(t,x,y) =& ~t^N - A_1\, t^{N-1} +t^{N-2}\left[A_2 - \binom{N}{2}\, m^2\,x\right]\\
&-t^{N-3}\left[A_3-\binom{N-1}{2}\,m^2\,A_1\,x  -\binom{N}{3} \, 2\,m^3\,y\right]\\
&+t^{N-4}\left[A_4-\binom{N-2}{2}\,m^2\,A_2\,x -\binom{N-1}{3}\, 2\,m^3\,A_1\,y\right.\\
&~~~~~\qquad\left.-\binom{N}{4}\,m^4\,(3\,x^2-\alpha\, E_4)\right] + O\big(t^{N-5}\big)=0~.
\end{aligned}
\label{F12}
\end{equation}
Since $F$ is a linear combination of the $P_n$, which are modular with weight $n$, it will
transform homogeneously (with weight $N$) if the coefficients $A_n$ are
modular with weight $n$. To verify this fact and provide a precise identification between the $A_n$ and the
the gauge invariant quantum observables $\vev{\Tr \Phi^n}$ which naturally parametrize the moduli space,
we find that the modifications that we have made to the $P_n$ as compared to those 
of \cite{Donagi:1995cf, Donagi:1997sr} are essential.

\subsection{The D'Hoker-Phong curve}
\label{secn:DPcurve}
The second form of the curve for the $\mathcal{N}=2^*$ U($N$) theory is due to D'Hoker and Phong
and was originally derived by using the relation between the SW curve 
and the spectral curve of the elliptic Calogero-Moser system \cite{D'Hoker:1997ha}. 
This spectral curve is abstractly defined as
\begin{equation}
R(t,z) \,\equiv\, \text{det} \Big[ t\,\one - L(z)\Big] = 0 \, ,
\label{DPcurve}
\end{equation}
where $L(z)$ is the Lax matrix of the integrable system. 
We refer the reader to \cite{D'Hoker:1997ha} for details and here we merely present the curve in the form
that is most convenient for our purposes. 

First, we define the degree $N$ polynomial $H(t)$:
\begin{equation}
H(t) = \prod_{u=1}^N(t-e_u) = \sum_{n=0}^N (-1)^n\,W_n\, t^{N-n}
\label{eq:defH}
\end{equation}
where 
\begin{equation}
W_n=\sum_{u_1< \cdots < u_n} e_{u_1}\cdots e_{u_n}~. 
\label{eq:defui}
\end{equation}
The $e_u$ are interpreted as the quantum-corrected vacuum expectation values of the 
scalar field $\Phi$ and, at weak coupling, they have the following form
\begin{equation}
e_u = a_u + O(q)
\end{equation}
in terms of the classical vacuum expectation values $a_u$ (see (\ref{au}). 
Thus, the gauge invariant quantum expectation values, which parametrize the quantum moduli space,
can be written as 
\begin{equation}
\vev{\Tr\Phi^n} = \sum_{u=1}^N \, e_u^n ~.
\label{TrPhi}
\end{equation}
Next, we define the function
\begin{equation}
f(t,z) = \sum_{n=0}^N (-1)^n\,\frac{m^n}{n!}\, h_n(z)\, H^{(n)}(t)
\label{eq:deff}
\end{equation}
where
\begin{equation}
H^{(n)}(t) \,\equiv\, \frac{d^nH(t)}{dt^n} = \sum_{\ell=0}^{N-n} (-1)^\ell\,\frac{(N-\ell)!}{(N-\ell-n)!} \,
W_\ell\, t^{N-n-\ell}~,
\label{eq:defHn}
\end{equation}
and
\begin{equation}
h_n(z) \,
\equiv\,\frac{1}{\theta_1(z\vert \tau)} \left(\frac{1}{2\pi \ii}\frac{d}{dz}\right)^n\, \theta_1(z \vert \tau)
\label{hn}
\end{equation}
with $\theta_1(z\vert \tau)$ being the first Jacobi $\theta$-function
\begin{equation}
\theta_1(z \vert \tau) = \sum_{n\in\mathbb{Z}}e^{\ii \pi \tau (n-\frac{1}{2})^2 + 2\pi \ii (z-\frac{1}{2}) 
(n-\frac{1}{2})} ~.
\end{equation}
Notice we have chosen normalizations so that the uniformizing coordinate $z$ on the 
torus obeys $z\sim z+1$ and $z\sim z+ \tau$, and that, as before, the complex 
structure parameter $\tau$ is identified with the gauge coupling (\ref{tau}).

Using this notation, the spectral curve of the Calogero-Moser system (\ref{DPcurve}),
and hence the SW curve for the U($N$) theory, takes the form \cite{D'Hoker:1997ha}
\begin{equation}
R(t,z) = f\big(t + m\,h_1, z\big) =0~.
\label{eq:Rf}
\end{equation}
To make the modular properties of the curve more manifest, we rewrite the 
function $f(t,z)$ in (\ref{eq:deff}) in a slightly different way.  We first observe that
\begin{equation}
  h_n(z)= \left(\frac{1}{2\pi\ii}\frac{d}{dz} + h_1(z)\right)^n 1~,
  \label{eq:hnrewrite}
\end{equation}
as one can easily check recursively. 
Plugging this into the definition \eqref{eq:deff} of $f$ and using (\ref{eq:defHn}) 
(and after a simple rearrangement of the sums),
we get
\begin{equation}
\begin{aligned}
  f(t,z)&=\sum_{n=0}^N \sum_{\ell=0}^{N-n} (-1)^{\ell+n}\binom{N-\ell}{n}\,
  W_\ell \,t^{N-\ell-n}\,m^n\,\left(\frac{1}{2\pi\ii}\frac{d}{dz} + h_1(z)\right)^n1 \\
  &=\sum_{\ell=0}^N (-1)^\ell\,W_\ell \left[ t - m \left( \frac{1}{2\pi\ii}\frac{d}{d z} + h_1(z) \right) \right]^{N-\ell} 1 ~.
\end{aligned}
\label{eq:frewrite}
\end{equation}
{From} this we see that the shift in $t$ in \eqref{eq:Rf} simply amounts to setting 
$h_1$=0 after taking the derivatives. Thus, the curve equation for the 
$\mathcal{N}=2^\star$ U($N$) theory in this formulation
becomes
\begin{equation}
\begin{aligned}
R(t,z)=&\sum_{\ell=0}^N (-1)^\ell\,W_\ell \left[ t - m \left( \frac{1}{2\pi\ii}
\frac{d}{d z} + h_1(z) \right) \right]^{N-\ell} 1~\Bigg\vert_{h_1=0} \\
=&~t^N -t^{N-1}\, W_1+ t^{N-2}\left[W_2 +\binom{N}{2}\,m^2\,h_1^\prime\right]\\
&-t^{N-3}\left[W_3 +\binom{N-1}{2}\,m^2\,h_1^\prime\,W_1
+\binom{N}{3}\,m^3\,h_1^{\prime\prime}\right]\\
&+t^{N-4}\left[W_4 +\binom{N-2}{2}\,m^2\,h_1^\prime\, W_2
+\binom{N-1}{3}\,m^3\,h_1^{\prime\prime}\,W_1\right.\\
&~~~~~\quad\quad\left.+\binom{N}{4}\,m^4\,
\big(h_1^{\prime\prime\prime}+3(h_1^\prime)^2\big)\right] + O(t^{N-5})=0 
\end{aligned}
\label{eq:Rrewrite}
\end{equation}
where the $^\prime$ stands for the derivative with respect to $2\pi\ii z$.

\subsection{Comparing curves}
\label{secn:curvecomp}
By comparing the two forms of the SW curve 
presented in the previous subsections, one can establish a relation between the $W_n$, 
which are related to the quantum expectation values $\vev{\Tr\Phi^n}$, 
and the modular covariant combinations $A_n$ on which S-duality acts in a 
simple way. A different method to relate the $A_n$ and the $W_n$, which only involves the 
D'€™Hoker-Phong form of the curve, is presented in Appendix~\ref{secn:modularDP}. 

Equating the coefficients of the same power of $t$ in (\ref{F12}) and (\ref{eq:Rrewrite}), we easily get
\begin{equation}
\begin{aligned}
A_1 &= W_1 ~,\\
A_2 &= W_2 +\binom{N}{2}\,m^2\,(h_1^\prime+x) ~,\\
A_3 &= W_3 +\binom{N-1}{2}\,m^2\,(h_1^\prime+x)\,W_1+\binom{N}{3}\,m^3\,(h_1^{\prime\prime}+2\,y)
~,\\
A_4 &=W_4+\binom{N-2}{2}\,m^2\,(h_1^\prime+x)W_2+\binom{N-1}{3}\,m^3\,(h_1^{\prime\prime}+2\,y)\\
&~\quad+
\binom{N}{4}\,m^4\,\big(h_1^{\prime\prime\prime}+3(h_1^\prime)^2+6\,h_1^\prime\,x+9\,x^2-\alpha\,E_4\big)
\end{aligned}
\label{AW}
\end{equation}
and so on. Recalling that $x$ and $y$ are related to the Weierstra\ss~function 
as shown in (\ref{xyvswp}), and using the properties of $\theta_1(z\vert\tau)$ 
and its derivatives, one can show that all $z$-dependence
cancels in the right hand side of (\ref{AW}) as it should, since
\begin{equation}
\begin{aligned}
&h_1^\prime+x = \frac{E_2}{12}~,\\
&h_1^{\prime\prime}+2\,y=0~,\\
&h_1^{\prime\prime\prime}+3(h_1^\prime)^2+6\,h_1^\prime\,x+9\,x^2
=\frac{E_2^2}{48}+\frac{E_4}{24}~.
\end{aligned}
\label{rel}
\end{equation}
We have included proofs of these identities in Appendix \ref{appeisen}. Using 
these results, the relations (\ref{AW}) simplify and reduce to
\begin{equation}
\begin{aligned}
A_1 &= W_1 ~,\\
A_2 &= W_2 +\binom{N}{2}\,\frac{m^2\,E_2}{12} ~,\\
A_3 &= W_3 +\binom{N-1}{2}\,\frac{m^2\,E_2}{12}\,W_1~,\\
A_4 &=W_4+\binom{N-2}{2}\,\frac{m^2\,E_2}{12}\,W_2+
\binom{N}{4}\,\Big(\frac{m^4\,E_2^2}{48}+\frac{m^4\,E_4(1-24\alpha)}{24}\Big)~.
\end{aligned}
\label{AW1}
\end{equation}
Notice that all terms proportional to $m^3$ cancel  and that the formula for 
$A_4$ can be further simplified by setting the free parameter to $\alpha=\frac{1}{24}$. With this choice we
eliminate the modular form $E_4$, leaving only the quasi-modular form $E_2$.

The same procedure may be carried out for the higher coefficients $A_n$ without 
any difficulty. Exploiting the
freedom of fixing the parameters in front of the modular forms to systematically eliminate them, 
we obtain the following rather compact result:
\begin{equation}
A_n = \sum_{\ell=0}^{\left[n/2\right]} \binom{N-n+2\ell}{2\ell}\,(2\ell-1)!! 
\left(\frac{m^2\,E_2}{12}\right)^\ell\, W_{n-2\ell} ~.
\label{eq:Akresult}
\end{equation} 
This formula can be easily inverted and one gets
\begin{equation}
\label{eq:Wkresult}
W_n = \sum_{\ell=0}^{\left[n/2\right]} (-1)^\ell\,\binom{N-n+2\ell}{2\ell} \,(2\ell-1)!! 
\left(\frac{m^2\,E_2}{12}\right)^\ell\, A_{n-2\ell} ~.
\end{equation}
We have verified these relations by working to higher orders in both $n$ and $N$. 
It is interesting to observe that, although both the Donagi-Witten curve and the 
D'Hoker-Phong curve separately have coefficients that are elliptic functions, the 
maps between the two sets of coefficients can be written entirely in terms of 
quasi-modular forms. For this to happen and, more importantly, in order 
that 
all dependence on the uniformizing coordinate $z$ disappears in the relations between the $A_n$ and the $W_n$,
it is essential to use a set of polynomials $P_n$ that are differ from those 
originally defined
in \cite{Donagi:1995cf, Donagi:1997sr}. 

Both $W_n$ and $A_n$ are good sets of coordinates for the Coulomb moduli space 
of the ${\mathcal N}=2^{\star}$ U($N$) SYM theory. The former naturally 
incorporate the quantum corrections that are calculable using either the curve analysis 
or by localization calculations while the latter are distinguished by their simple 
behavior under S-duality. In the following sections, we will independently calculate 
the $A_n$ and the $W_n$ in a weak-coupling expansion and show that they satisfy the general relations
\eqref{eq:Akresult} and (\ref{eq:Wkresult}) provided some important caveats are taken into account.

\section{Period integrals and modular anomaly equation}
\label{secn:periodsandmodularity}

In this section, we present two methods to compute the modular covariant quantities $A_n$ and express them in terms
of the classical vacuum expectation values $a_u$ of the adjoint scalar field $\Phi$ given in (\ref{au}).
The first method is based on a direct use of the curve and the associated differential, while 
the second exploits an extension of the modular anomaly equation (\ref{recdiff}).

\subsection{Period integrals}
By solving the Donagi-Witten curve equation (\ref{DWcurve0}) one can express the 
variable $t$ as a function of $x$ and $y$, and hence of the uniformizing 
coordinate of the torus $z$ through the identifications (\ref{xyvswp}). 
Once this is done, the SW differential is given by \cite{D'Hoker:1997ha}:
\begin{equation}
\lambda_{\mathrm{SW}} = t(z)\, dz ~,
\end{equation}
and its periods are identified with the pairs of dual variables $a_u$ and $a_u^{\mathrm{D}}$ according to
(\ref{vevperiods}). Of course, in order to obtain explicit expressions, a 
canonical basis of $1$-cycles is needed. Since the curve is an $N$-fold cover 
of a torus, there is a natural choice for such a basis, as we now demonstrate.
In fact, $F$ being a polynomial of degree $N$, we can factorize it as
\begin{equation}
F = \prod_{u=1}^N \Big(t-t_{u}(x(z),y(z))\Big) = 0~,
\end{equation}
and then define
\begin{equation}
\label{Aperiod}
\begin{aligned}
a_u &= \oint_{A_u}\lambda_{\mathrm{SW}}  := \oint_{\alpha} t_{u}\big(x(z),y(z)\big)\,dz~,\\
a_u^{\mathrm{D}} &= \oint_{B_u}\lambda_{\mathrm{SW}}  := \oint_{\beta} t_{u}\big(x(z),y(z)\big)\,dz~,\\
\end{aligned}
\end{equation}
where $\alpha$ and $\beta$ are, respectively, the $A$ and $B$ cycles of the torus. 
To see that this identification is correct, let us (for a moment) consider switching 
off the mass of the adjoint hypermultiplet.
If we do so, the supersymmetry is enhanced to $\mathcal{N}=4$ and Donagi-Witten polynomials 
simply become $P_n=t^n$, so that the curve takes the form
\begin{equation}
\label{Fmasslessv1}
F = \sum_{n=0}^N (-1)^{n} t^{N-n}\, A_{n} =0~.
\end{equation}
Since in the ${\mathcal N}=4$ SYM theory the classical moduli space does not receive quantum corrections,
it makes sense to identify the modular covariant coordinates $A_n$ with the symmetric 
polynomials constructed from the classical vacuum expectation values, namely
\begin{equation}
\label{Akzero}
A_n = \sum_{{u_1} < \cdots < u_{n}} a_{u_1}\cdots a_{u_n} ~.
\end{equation}
Substituting this into \eqref{Fmasslessv1}, we see that $F$ factorizes as
\begin{equation}
\label{Fmasslessv2}
F = \prod_{u=1}^N (t-a_u) = 0 \, ,
\end{equation}
so we may conclude that in the massless limit we have $t_{u}=a_u$. This is clearly 
consistent with our ansatz (\ref{Aperiod}), since the integral over the $\alpha$-cycle 
gives unity. The integral over the $\beta$-cycle, instead, gives
\begin{equation}
a^{\text{D}}_u = \oint_{\beta} a_u = \tau\, a_u ~,
\end{equation}
which is the expected answer in the $\mathcal{N}=4$ gauge theory. 

Let us now revert to our original problem, and consider the scenario where the adjoint 
hypermultiplet has a mass $m$. In general, it is not possible to compute 
the period integrals (\ref{Aperiod}) explicitly, as each of the $t_{u}(x,y)$ is a solution of
a generic polynomial equation of degree $N$.  However, progress can be made by assuming that 
each of these solutions has a expansion in powers of the hypermultiplet mass, of the form
\begin{equation}
t_{u}(x,y) = a_u + \sum_{\ell\in\,\mathbb{N}/2} t_{u}^{(\ell)}(x,y)\,m^{2\ell}~,
\label{texp}
\end{equation}
and by working perturbatively order by order in $m$. Notice that in (\ref{texp}) the sum is over both 
integers and half-integers in order to have in principle both even and odd powers of $m$, even though in the
end only the even ones will survive. Of course, this assumption implies that
the modular covariant coordinates on moduli space have a mass expansion of the form
\begin{equation}
A_n = \sum_{{u_1} < \cdots < u_{n}} a_{u_1}\cdots a_{u_n} + \sum_{\ell\in\,\mathbb{N}/2}  
A_{n}^{(\ell)}\,m^{2\ell} ~.
\label{Aexp}
\end{equation}
Using this ansatz in the curve equation  \eqref{F12} leads to constraints on the 
$t_{u}^{(\ell)}$, which we solve in terms of the $A_{n}^{(\ell)}$. Finally, we 
substitute these into the expressions for the 
$A$-periods in \eqref{Aperiod} and demand that all higher order terms in $m$ vanish for self-consistency
as that equation is already solved by $t_u^{(0)}$. 
The integrals for these higher order terms typically involve integrals of powers 
of the Weierstra\ss~function and
its derivative, which are known in terms of quasi-modular forms. In this way we 
can construct the various mass corrections
$A_{n}^{(\ell)}$ in terms of the classical $a_u$ and of quasi-modular forms. 

Let us first illustrate this procedure in the simple case of the U$(2)$ gauge theory.
For $N=2$ the Donagi-Witten curve is
\begin{equation}
t^2 -t A_1 + (A_2 - m^2 x) = 0~.
\end{equation}
Inserting the mass expansions (\ref{texp}) and (\ref{Aexp}) and collecting the powers of $m$, we obtain
\begin{align}
&a_u^2-a_u (a_1+a_2)+a_1\,a_2
+m \left(A_{2}^{(1/2)}+(2 \,a_u-a_1-a_2)\, t^{(1/2)}_u-a_u\,A_{1}^{(1/2)}\right) \label{curveU2exp}\\
&+m^2 \left(A_{2}^{(1)}+\big(t^{(1/2)}_u\big)^2+(2\,a_u-a_1-a_2) 
\,t^{(1)}_u-t^{(1/2)}_u A_{1}^{(1/2)}-a_u\, A_{1}^{(1)}-x\right) +O(m^3)= 0\notag
\end{align}
for $u=1,2$. It is easy to check that the zeroth order term in the mass vanishes, as it should.
Requiring the cancellation of the term at linear order in $m$ amounts to setting
\begin{equation}
t_u^{(1/2)} = \frac{A_{2}^{(1/2)}-a_u\, A_{1}^{(1/2)}}{a_1+a_2-2 a_u}
\label{tu1}
\end{equation}
for $u=1,2$. Now, in order to maintain the relation \eqref{Aperiod}, the 
integral of $t_u^{(\ell)}$ over the $A$-cycles has to vanish for all $\ell$. 
In particular, for $\ell=1/2$ and taking into account that $t_u^{(1/2)}$ in (\ref{tu1}) 
is constant with respect to $z$, one has
\begin{equation}
\oint_\alpha t_u^{(1/2)}\,dz = \frac{A_{2}^{(1/2)}-a_u\, A_{1}^{(1/2)}}{a_1+a_2-2 a_u} = 0
\end{equation}
for both $u=1$ and $u=2$. In turn this leads to
\begin{equation}
A_{1}^{(1/2)}=A_{2}^{(1/2)} = 0 ~.
\end{equation}
Substituting this into (\ref{curveU2exp}) and demanding the cancellation of the $m^2$ terms, we get
\begin{equation}
t^{(1)}_u= \frac{A_{2}^{(1)}-a_u A_{1}^{(1)}-x}{a_1+a_2-2 a_u}~.
\end{equation}
Imposing that
\begin{equation}
\oint_\alpha t_u^{(1)}\,dz = 0
\end{equation}
for $u=1,2$, and using the fact that, in view of the identification (\ref{xyvswp}),
\begin{equation}
\oint_{\alpha} x\,dz =\frac{1}{(2\pi\ii)^2}\oint_{\alpha}\wp(z)\,dz= \frac{E_2}{12}~,
\label{intalpha}
\end{equation}
we get
\begin{equation}
A_{1}^{(1)} = 0~, \qquad A_{2}^{(1)} = \frac{E_2}{12} ~.
\end{equation}
Recapitulating, we have obtained 
\begin{equation}
\begin{aligned}
A_1 &= a_1+a_2~,\\
A_2&=a_1a_2 + \frac{m^2}{12}\,E_2 + O(m^3)~.
\end{aligned}
\end{equation}
This process can be repeated in similar fashion to obtain all mass corrections in 
a systematic way.  This procedure requires that we compute period integrals of 
polynomials in the Weirstra\ss~function and its derivative which can be done using 
standard techniques (see for example \cite{KashaniPoor:2012wb} and references therein). 
We stress that although this approach is perturbative in $m$, it is \emph{exact} in the 
gauge coupling constant, since the coefficients 
are fully resummed quasi-modular forms in $\tau$. 

The same procedure can of course be carried out for $\mathcal{N}=2^\star$ theories with higher 
rank gauge groups, even if the calculations quickly become more involved as $N$ increases.
The results, however, can be organized in a rather compact way by using the
lattice sums $C^{\,p}_{n;m_1\cdots}$ defined in (\ref{Cpnm}). In fact, 
the expressions we find for the first few $A_n$ at the first few 
non-trivial orders in $m$ in the U($N$) theory are
\begin{align}
A_1 &=\sum_{u} a_u ~,\phantom{\Bigg|}\label{A1UN}\\
A_2 &= \sum_{u_1<u_2} a_{u_1} a_{u_2} 
+\binom{N}{2}\, \frac{m^2}{12}\,E_2+ \frac{m^4}{288}\,\big(E_2^2-E_4\big)\,C^{\,0}_{\,2}
+\frac{m^6}{4320}\,\big(5E_2^3-3E_2E_4-2E_6\big)\,C^{\,0}_{\,4}\notag\\
&\quad\quad+\frac{m^6}{3456}\,\big(E_2^3-3E_2E_4+2E_6\big)\,C^{\,0}_{\,2;11} + O(m^8)~,\phantom{\Bigg|}
 \label{A2UN}\\
A_3 &= \!\!\sum_{u_1<u_2<u_3} a_{u_1} a_{u_2}a_{u_3}
+\binom{N-1}{2}\,\frac{m^2}{12}\, E_2 \sum_{u}a_u+\frac{m^4}{288}\,\big(E_2^2-E_4\big)
\Big(C^{\,0}_{\,2}\,\sum_{u} a_u
- 2\, C^{\,1}_{\,2} \Big)\notag\\
&\quad\quad +\frac{m^6}{4320}\,\big(5E_2^3-3E_2E_4-2E_6\big)\Big(C^{\,0}_{\,4} 
\,\sum_{u} a_u-2\, C^{\,1}_{\,4}\Big)\notag\\
&\quad\quad +\frac{m^6}{3456}\,\big(E_2^3-3E_2E_4+2E_6\big)\Big(C^{\,0}_{\,2;11} 
\,\sum_{u} a_u-2\, C^{\,1}_{\,2;11}\Big)+O(m^8)~,\phantom{\Bigg|}\label{A3UN}\\
A_4 &= \!\!\sum_{u_1<\cdots<u_4} a_{u_1} a_{u_2}a_{u_3}a_{u_4}+\binom{N-2}{2}
\,\frac{m^2}{12}\,E_2 \sum_{u_1<u_2} a_{u_1} a_{u_2} +\binom{N}{4}\,\frac{m^4}{48}\,E_2^2
\notag\\
&\quad\quad+\frac{m^4}{288}\,(E_2^2-E_4)\Big(C^{\,0}_{\,2} \sum_{u_1<u_2} a_{u_1} a_{u_2} -
2\,C^{\,1}_{\,2} \,\sum_u a_u + 3\,C^{\,2}_{\,2} -\binom{N}{2}\Big) \notag\\
&\quad\quad +\frac{m^6}{4320}\,\big(5E_2^3-3E_2E_4-2E_6\big)
\Big(C^{\,0}_{\,4}\sum_{u_1<u_2} a_{u_1} a_{u_2} -2\, C^{\,1}_{\,4}\,\sum_{u} a_u+3\,C^{\,2}_{\,4} 
-\frac{1}{2}\,C^{\,0}_{\,2}\Big)
\notag\\
&\quad\quad +\frac{m^6}{3456}\,\big(E_2^3-3E_2E_4+2E_6\big)\Big(C^{\,0}_{\,2;11}
\sum_{u_1<u_2} a_{u_1} a_{u_2} -2\, C^{\,1}_{\,2;11}\,\sum_{u} a_u+3\,C^{\,2}_{\,2;11} 
\Big)\notag\\
&\quad\quad +\binom{N-2}{2}\,\frac{m^6}{3456}\,E_2\big(E_2^2-E_4\big)\,C^{\,0}_{\,2} + O(m^8) ~.
\phantom{\Bigg|}
\label{A4UN}
\end{align}
Of course, only the $A_n$ with $n\le N$ are the independent coordinates that can 
be used to parametrize the moduli space of the theory.
Despite their appearance, it is not difficult to recognize a regular pattern in 
these expressions, which contain the
same combinations of Eisenstein series appearing in
the prepotential coefficients.
Notice also that only even powers of $m$ are present, this being in full agreement 
with the $\mathbb{Z}_2$ symmetry of the theory that sends $m\to-m$. 

We have explicitly verified that under S-duality the above $A_n$ transform with weight $n$, namely
\begin{equation}
S(A_n)= \tau^n\,A_n~.
\label{SonA}
\end{equation}
To do so we used the properties of the Eisenstein series under inversion, and replaced
each $a_u$ with the corresponding dual variable $a_u^{\mathrm{D}}$, which can be computed either by
evaluating the periods of the SW differential along the $B$-cycles according to 
(\ref{Aperiod}) or, more efficiently, by taking the derivative of the prepotential 
with respect to $a_u$ according to (\ref{adual}).
The fact that (\ref{SonA}) holds true despite the explicit presence of the quasi-modular 
Eisenstein series $E_2$ in the $A_n$ is a highly non-trivial consistency check.
Finally, we observe that by inserting (\ref{A1UN})--(\ref{A4UN}) in the map (\ref{eq:Wkresult}), 
one can obtain the quantum expectation values 
$W_n$ in terms of the classical variables $a_u$. The result is
\begin{align}
W_1 &=\sum_{u} a_u ~,\label{W1q}\\
W_2 &= \sum_{u_1<u_2} a_{u_1} a_{u_2} 
+ \frac{m^4}{288}\,\big(E_2^2-E_4\big)\,C^{\,0}_{\,2}
+\frac{m^6}{4320}\,\big(5E_2^3-3E_2E_4-2E_6\big)\,C^{\,0}_{\,4}\notag\\
&\quad\quad+\frac{m^6}{3456}\,\big(E_2^3-3E_2E_4+2E_6\big)\,C^{\,0}_{\,2;11} + O(m^8)~,
 \label{W2q}\\
W_3 &= \!\!\sum_{u_1<u_2<u_3} a_{u_1} a_{u_2}a_{u_3}
+\frac{m^4}{288}\,\big(E_2^2-E_4\big)\Big(C^{\,0}_{\,2}\,\sum_{u} a_u
- 2\, C^{\,1}_{\,2} \Big)\notag\\
&\quad\quad +\frac{m^6}{4320}\,\big(5E_2^3-3E_2E_4-2E_6\big)\Big(C^{\,0}_{\,4} 
\,\sum_{u} a_u-2\, C^{\,1}_{\,4}\Big)\notag\\
&\quad\quad +\frac{m^6}{3456}\,\big(E_2^3-3E_2E_4+2E_6\big)\Big(C^{\,0}_{\,2;11} 
\,\sum_{u} a_u-2\, C^{\,1}_{\,2;11}\Big)+O(m^8)~,\label{W3q}\\
W_4 &= \!\!\sum_{u_1<\cdots<u_4} a_{u_1} a_{u_2}a_{u_3}a_{u_4} 
\notag\\
&\quad\quad+\frac{m^4}{288}\,(E_2^2-E_4)\Big(C^{\,0}_{\,2} \sum_{u_1<u_2} a_{u_1} a_{u_2} -
2\,C^{\,1}_{\,2} \,\sum_u a_u + 3\,C^{\,2}_{\,2} -\binom{N}{2}\Big) \notag\\
&\quad\quad +\frac{m^6}{4320}\,\big(5E_2^3-3E_2E_4-2E_6\big)
\Big(C^{\,0}_{\,4}\sum_{u_1<u_2} a_{u_1} a_{u_2} -2\, C^{\,1}_{\,4}\,\sum_{u} a_u+3\,C^{\,2}_{\,4} 
-\frac{1}{2}\,C^{\,0}_{\,2}\Big)
\notag\\
&\quad\quad +\frac{m^6}{3456}\,\big(E_2^3-3E_2E_4+2E_6\big)
\Big(C^{\,0}_{\,2;11}\sum_{u_1<u_2} a_{u_1} a_{u_2} -2\, C^{\,1}_{\,2;11}\,\sum_{u} a_u+3\,C^{\,2}_{\,2;11} 
\Big)\notag\\
&\quad\quad + O(m^8) ~.
\label{W4q}
\end{align}
It is interesting to notice that these expressions are a bit simpler than the ones for 
the $A_n$; in particular, all
$m^2$ terms disappear and, up to a constant term in $W_4$, all other explicit dependence on 
$N$ drops out. These formulas will be useful in later sections, where we compare them with 
results from explicit localization calculations.
An important consistency check on our results is the fact that both $W_3$ and $W_4$ vanish for U(2),
and that $W_4$ vanishes for U(3). This has to happen since the $W_n$ are symmetric polynomials 
in the quantum variables $e_u$, see (\ref{eq:defui}).

\subsection{Modular anomaly equation}

We now explore an alternative route to express the $A_n$ in terms of the classical parameters $a_u$,
which is based on the S-duality transformation properties. The main idea is simple: if we assume the mass
expansion (\ref{Aexp}), 
then the requirement that $A_n$ transforms with weight $n$ under S-duality
constrains the form of $A_n^{(\ell)}$ once the previous mass terms are known.
So, starting from the classical part it is possible to systematically reconstruct in this way all subleading terms.

Let us recall from Section \ref{secn:rev} that\,%
\footnote{For simplicity we suppress the subscripts and denote the pair 
$(a_u,a_u^{\text{D}})$ as $(a,a^{\text{D}})$.}
\begin{equation}
S(a) = a^{\text{D}}= \frac{1}{2\pi \ii}\, \frac{\partial F}{\partial a}=
 \tau\left(a + \frac{\delta}{12} \frac{\partial f}{\partial a} \right)
\label{Sona1}
\end{equation}
where $f$ is the quantum part of the prepotential and $\delta = \frac{6}{\ii\pi\tau}$. 
Furthermore, in order for the $A_n$ to have the correct mass dimension, the subleading terms 
$A_{n}^{(\ell)}$ must be homogeneous functions of $a$ with weight $n-2\ell$:
\begin{equation}
A_{n}^{(\ell)}(\tau, \lambda \,a) = \lambda^{n-2\ell}\, A_{n}^{(\ell)}(\tau, a)~.
\end{equation}
The other basic requirement is that they are quasi modular forms of weight $2\ell$. This implies that the
$A_{n}^{(\ell)}$ depend on the coupling constant $\tau$ only through the 
Eisenstein series $E_2$, $E_4$ and $E_6$, namely
\begin{equation}
A_{n}^{(\ell)}(\tau,a)= A_{n}^{(\ell)}\big(E_2(\tau),E_4(\tau),E_6(\tau),a\big) ~,
\end{equation}
so that
\begin{equation}
\begin{aligned}
A_{n}^{(\ell)}\big(\!-\ft{1}{\tau},a\big)& =
A_{n}^{(\ell)}
\Big(E_2\big(\!-\ft{1}{\tau}\big),E_4\big(\!-\ft{1}{\tau}\big),
E_6\big(\!-\ft{1}{\tau}\big),a\Big) \\ &=\tau^{2\ell}\,A_{n}^{(\ell)}\big(E_2+\delta,E_4,E_6,a\big) ~,
\end{aligned}
\end{equation}
where in the last step we have used the anomalous modular transformation (\ref{E2s}) of the second Eisenstein series
$E_2$. From now on, for ease of notation, we only exhibit the dependence on $E_2$. 
Putting everything together, we find
\begin{equation}
\begin{aligned}
S\big(A_{n}^{(\ell)}\big)& = A_{n}^{(\ell)}
\Big(E_2\big(\!-\ft{1}{\tau}\big), a^{\text{D}}\Big) 
= \tau^{n}\, A_{n}^{(\ell)}\Big(E_2+\delta, a+\ft{\delta}{12}\,\ft{\partial f}{\partial a} \Big)\\
&=\tau^{n}\, \left[A_{n}^{(\ell)}+
\Big(\frac{\partial A_{n}^{(\ell)}}{\partial E_2} + \frac{1}{12} \frac{\partial A_{n}^{(\ell)}}{\partial a}
\cdot \frac{\partial f}{\partial a}\Big)\,\delta+O\big(\delta^2\big)\right]~.
\end{aligned}
\label{SonAnl}
\end{equation}
The requirement that under S-duality $A_n$ be a modular form of weight $n$ leads 
to a modular anomaly equation:
\begin{equation}
\frac{\partial A_n}{\partial E_2} + \frac{1}{12} \frac{\partial A_n}{\partial a}
\cdot \frac{\partial f}{\partial a} = 0~.
\label{modeqA}
\end{equation}
Notice that if (\ref{modeqA}) is satisfied, then all terms in (\ref{SonAnl}) which 
are of higher order in $\delta$, vanish.
Expanding both the $A_n$ and the quantum prepotential $f$ in powers of $m$, 
we can rewrite the above modular anomaly equation in the form of a recursion relation for the 
$A_{n}^{(\ell)}$,
namely
\begin{equation}
\frac{\partial A_{n}^{(\ell)}}{\partial E_2} + 
\frac{1}{12} \sum_{k=0}^{\ell}\frac{\partial A_{n}^{(k)}}{\partial a}\cdot 
\frac{\partial f_{\ell-k}}{\partial a} = 0~.
\label{recA}
\end{equation}
This shows that starting from the classical symmetric polynomials
\begin{equation}
A_{n}^{(0)}= \sum_{u_1<\cdots u_n} a_{u_1}\cdots a_{u_n}
\label{An0}
\end{equation}
and the prepotential coefficients (some of which have been listed in (\ref{fnexpl})), one can systematically
calculate the higher order terms and obtain the modular completion iteratively by integrating the
modular anomaly equation (\ref{recA}). For example, at the first step ($\ell=1$) we have
\begin{equation}
\begin{aligned}
\frac{\partial A_{n}^{(1)}}{\partial E_2} =- \frac{1}{12} 
\frac{\partial A_{n}^{(0)}}{\partial a}\cdot \frac{\partial f_{1}}{\partial a} 
= -\frac{1}{12} \sum_{u\ne v}\frac{\partial A_{n}^{(0)}}{\partial a_u}\, \frac{1}{a_u-a_v}~,
\end{aligned}
\end{equation}
which is solved by  
\begin{equation}
A_{n}^{(1)}=\binom{N-n+2}{2} \,\frac{E_2}{12}\, A_{n-2}^{(0)}~.
\end{equation}
The higher order corrections $A_{n}^{(\ell)}$ can be similarly derived up to terms that are purely 
composed of modular forms of weight $2\ell$. These cannot be determined from the recursion 
relation alone, which is a symmetry requirement, and some extra dynamical input is needed.  
To illustrate this point let us consider the explicit expressions of $A_1$ and $A_2$ for the U($N$) theory
that can be derived using the above procedure. Up to order $m^8$ we find
\begin{align}
A_1 &=\sum_{u} a_u ~,\\
A_2 &= \sum_{u_1<u_2} a_{u_1} a_{u_2} 
+\binom{N}{2}\,\frac{m^2}{12}\, E_2+ \frac{m^4}{288}\,\big(E_2^2-\alpha\,E_4\big)\,C^{\,0}_{\,2}\notag \\
&\quad\quad+\frac{m^6}{4320}\,\big(5E_2^3+(2-5\alpha)E_2E_4-\beta\,E_6\big)\,C^{\,0}_{\,4} \notag\\
&\quad\quad+\frac{m^6}{3456}\,\big(E_2^3-(2+\alpha)E_2E_4+\gamma\,E_6\big)\,C^{\,0}_{\,2;11} + O(m^8)~,\label{A12UN}
\end{align}
where $\alpha,\beta,\gamma$ are free parameters. As anticipated, the terms that only depend on $E_2$ are
completely fixed by the modular anomaly equation, while those involving also the modular forms $E_4$ and
$E_6$ depend on integration constants. 
One can fix them by requiring that the perturbative limit of the above expressions, in which 
all Eisenstein series effectively are set to $1$, matches with the known perturbative behavior 
that can be deduced from the relations between
the modular $A_n$ and the quantum $W_n$ discussed in Section~\ref{secn:curvecomp}. In particular,
from (\ref{eq:Akresult}) with $n=2$ we see that
\begin{equation}
A_2\big|_{\text{cl}} = W_2\big|_{\text{cl}}+\binom{N}{2}\, \frac{m^2}{12}=
 \sum_{u_1<u_2}a_{u_1}a_{u_2} +\binom{N}{2}\, \frac{m^2}{12}~.
\end{equation}
This perturbative behavior is matched by (\ref{A12UN}) only if 
\begin{equation}
\alpha=1\quad\text{and}\quad\beta=\gamma=2~.
\end{equation}
It is reassuring to see that with this choice of parameters one precisely recovers the expression
for $A_2$ in (\ref{A2UN}) that was obtained from the calculation of the period integrals. 
By extending this procedure to higher order we can also derive $A_3$ and $A_4$ and verify that they
exactly agree with (\ref{A3UN}) and (\ref{A4UN}). This match is a very
strong indication of the correctness of our calculations and the validity of the approach based on the
modular anomaly equation (\ref{modeqA}).

Finally, we would like to remark that up to order $m^{10}$ the matching with the perturbative results is enough
to completely fix all integration constants, since there is a unique modular form of weight $2n$ up to
$n=5$. At $n=6$, i.e.~at order $m^{12}$ there are two independent modular forms of weight
12, namely $E_4^3$ and $E_6^2$. So the knowledge of the perturbative behavior is not enough to fix
all parameters and more information, for example from the 1-instanton sector, is needed. 
At $n=7$, again the perturbative information is sufficient since only one modular form of weight 14 exists.
However from that point on, some extra data from the non-perturbative sectors is necessary.
This is exactly the same situation occurring also for the prepotential coefficients, as
pointed out for instance in \cite{Billo':2015ria,Billo':2015jta,Billo:2016zbf}.

\section{Chiral observables from localization}
\label{secn:localization}

The discussion of the previous section clearly shows that in order to confirm the general relations among the
chiral observables and their modular properties, and also to have data to fix the coefficients left undetermined by
the modular anomaly equation, it is necessary to explicitly compute some instanton contributions. 
This is possible using the equivariant localization techniques.

Following the discussion in \cite{Billo':2015ria}, we first deform the $\mathcal{N}=2^\star$ 
theory by introducing the $\Omega$-background \cite{Nekrasov:2002qd,Nekrasov:2003rj} and 
then calculate the partition function in a multi-instanton sector. The $\Omega$-deformation parameters 
will be denoted $\epsilon_1$ and $\epsilon_2$. The partition function $Z_k$ for the U$(N)$ theory 
in the presence of $k$-instantons is obtained by doing the following multi-dimensional contour integral:
\begin{equation}
Z_k = \oint \prod_{i=1}^k \frac{d\chi_i}{2\pi \ii}\, z_k^{\text{gauge}}\, z_k^{\text{matter}}~,
\label{Zk}
\end{equation}
where the integrand is given by
\begin{subequations}
\label{integrands}
\begin{align}
z_k^{\text{gauge}} &= \frac{(-1)^k}{k!}\left(\frac{\epsilon_1+\epsilon_2}{\epsilon_1\epsilon_2} 
\right)^k \frac{\Delta(0)\Delta(\epsilon_1+\epsilon_2)}{\Delta(\epsilon_1)\Delta(\epsilon_2)}
\prod_{i=1}^k \frac{1}{P(\chi_i+\frac{\epsilon_1+\epsilon_2}{2})P(\chi_i-\frac{\epsilon_1+\epsilon_2}{2})}\\
z_k^{\text{matter}}&=\left(\frac{(\epsilon_1+\epsilon_3)(\epsilon_1+\epsilon_4)}{\epsilon_3\epsilon_4} 
\right)^k \frac{\Delta(\epsilon_1+\epsilon_3)\Delta(\epsilon_1+\epsilon_4)}{\Delta(\epsilon_3)
\Delta(\epsilon_4)}\prod_{i=1}^k P(\chi_i+\ft{\epsilon_3-\epsilon_4}{2})P(\chi_i-\ft{\epsilon_3-\epsilon_4}{2})
\end{align}
\end{subequations}
with
\begin{equation}
P(x) = \prod_{u=1}^N (x-a_u) \qquad \Delta(x) = \prod_{i<j}^k(x^2- \chi_{ij}^2) ~,
\end{equation}
and $\chi_{ij} = \chi_i - \chi_j$. 
The parameters $\epsilon_3$ and $\epsilon_4$ are related the hypermultiplet mass $m$ according to
\begin{equation}
\epsilon_3 = m - \frac{\epsilon_1+\epsilon_2}{2}~,\qquad \epsilon_4 = -m -\frac{\epsilon_1+\epsilon_2}{2} ~.
\end{equation}
The contour integrals are computed by closing the contours in the upper half planes of the $\chi_i$ 
variables, assigning imaginary parts to the $\epsilon$'s, with the prescription~\cite{Billo':2015ria}:
\begin{equation}
\label{contourpresc}
\mathrm{Im}(\epsilon_4) \gg \mathrm{Im}(\epsilon_3) \gg \mathrm{Im}(\epsilon_2) \gg 
\mathrm{Im}(\epsilon_1) >0 ~.
\end{equation}
This prescription allows one to calculate the residues without ambiguity and obtain the partition function 
\begin{equation}
Z_{\text{inst}} = 1 + \sum_k q^k Z_k\,,
\end{equation}
from which one can derive the instanton part of prepotential
\begin{equation}
F_{\text{inst}} = \lim_{\epsilon_1,\epsilon_2\to 0}\Big(\!\!-\epsilon_1\epsilon_2 \log Z_{\text{inst}} 
\Big) = \sum_{k=1} q^k\, F_k ~.
\end{equation}
In this way one can compute the non-perturbative contributions to the coefficients $f_n$ and verify the
agreement with the resummed expressions like those given in (\ref{fnexpl}) (for details we refer to 
\cite{Billo':2015ria,Billo:2012st} and references therein).

The same localization methods can be used to compute the chiral correlators, which are known to receive 
quantum corrections from all instanton sectors. In this framework the expectation value for the 
generating function of such chiral observables is given by
\cite{Bruzzo:2002xf, Losev:2003py, Flume:2004rp,Billo:2012st}
\begin{equation}
\begin{aligned}
\vev{\Tr \rme^{z\,\Phi}}\big|_{\text{loc}}&=\sum_{n=0}\frac{z^n}{n!}\,\vev{\Tr \Phi^n}\big|_{\text{loc}}\\
&=
\sum_{u=1}^N e^{z a_u} -\frac{1}{Z_{\text{inst}}}
\sum_{k=1}^{\infty} \frac{q^k}{k!}\oint \prod_{i=1}^k \frac{d\chi_i}{2\pi \ii} \, 
{\mathcal O}(z, \chi_i)\,z_k^{\text{gauge}}\, z_k^{\text{matter}}~,
\end{aligned}
\label{trphin1}
\end{equation}
where the operator insertion in the instanton partition function is explicitly given by
\begin{equation}
{\mathcal O}(z, \chi_i) = \sum_{i=1}^k \rme^{z\chi_i}(1-e^{z\epsilon_1})(1-e^{z\epsilon_2})~,
\end{equation}
and the prescription to perform the contour integrals in \eqref{contourpresc} is the same as the one 
used for the instanton partition function. By explicitly computing these integrals order by order in $k$ and then
taking multiple derivatives with respect to $z$, one obtains the various instanton contributions to
the chiral observables $\vev{\Tr \Phi^n}\big|_{\text{loc}}$. 
Up to three instantons and for $n\le 5$, we have explicitly 
verified that these instanton corrections can be compactly written using the lattice sums (\ref{Cpnm}) as follows
\begin{align}
\vev{\Tr \Phi^n}\big|_{\text{loc}}&\!\!= C^{\,n}
-\binom{n}{2}2\,m^2(q+3q^2+4q^3+\cdots)\,C^{\,n-2}\notag\\
&~~+\binom{n}{2}2\,m^4(q+6q^2+12q^3+\cdots)\,C^{\,n-2}_{\,2}+
\binom{n}{4}2\,m^4(3q^2+20q^3+\cdots)\,C^{\,n-4}\notag\\
&~~-\binom{n}{2}24\,m^6(q^2+8q^3+\cdots)\,C^{\,n-2}_{\,4}
+\binom{n}{2}\,m^6(q+12q^2+36q^3+\cdots)\,C^{\,n-2}_{\,2;11}\notag\\
&~~-\binom{n}{4}24\,m^6(q^3+\cdots)\,C^{\,n-4}_{\,2}+O(m^8)~.
\label{Trphinq}
\end{align}
Recall that $C^{\,n}=\sum_{u}a_u^n$ and that one should set the $C$s to zero when the superscript 
of the $C$'s is negative. Based on our previous experience we expect that the coefficients 
of the various structures in (\ref{Trphinq}) are just the first terms of the instanton 
expansion of (quasi)-modular forms built out of Eisenstein series.
This is indeed what happens. In fact, we find
\begin{align}
\vev{\Tr \Phi^n}\big|_{\text{loc}}&\!\!= C^{\,n}
+\binom{n}{2}\frac{m^2}{12}(E_2-1)C^{\,n-2}
\notag\\
&-\binom{n}{2}\frac{m^4}{144}\big(E_2^2-E_4\big)C^{\,n-2}_{\,2}+
\binom{n}{4}\frac{m^4}{720}\big(21-30E_2+10E_2^2-E_4\big)C^{\,n-4}\notag\\
&-\binom{n}{2}\frac{m^6}{2160}\big(5 E_2^3 - 3 E_2 E_4 - 2 E_6\big)C^{\,n-2}_{\,4}
 - \binom{n}{2}\frac{m^6}{1728}\big(E_2^3 - 3 E_2 E_4 + 2 E_6\big)C^{\,n-2}_{2;11}  \notag\\
&+ \binom{n}{4}\frac{m^6}{4320}\big(15 E_2^2 - 5 E_2^3 - 15 E_4 + 9 E_2 E_4 - 4 E_6\big)
C^{\,n-4}_{\,2}
 +O(m^8)~.
 \label{Trphin}
\end{align}
By expanding the Eisenstein series in powers of $q$ we can obtain the contributions at 
any instanton number. We have verified the correctness of our extrapolation by computing the 4 and 5 instanton terms
in the U(4) theory and the 4 instanton terms in the U(5) theory, finding perfect match with the ``predictions''
coming from the Fourier expansion of (\ref{Trphin}). We also note that using the Matone relation
\cite{Matone:1995rx}, the result for $n=2$ matches perfectly with the mass expansion 
of the prepotential obtained in \cite{Billo':2015ria,Billo':2015jta}. 
Another noteworthy feature of the formula (\ref{Trphin}) 
is that the same quasi-modular functions appear for all values of $n$. Our results can therefore be 
thought of as a natural generalization of the result for the prepotential to other observables of
the gauge theory.

To compare with our findings of the previous sections,
it is convenient to change basis and make combinations of the above operators 
that describe the quantum version of the symmetric polynomials in the classical vacuum 
expectation values. At the first few levels the explicit map is
\begin{equation}
\begin{aligned}
W_1^{\text{loc}}&=\vev{\Tr \Phi}\big|_{\text{loc}}~,\\
W_2^{\text{loc}}&=\frac{1}{2}\Big(\vev{\Tr \Phi}\big|_{\text{loc}}^2-\vev{\Tr \Phi^2}\big|_{\text{loc}}\Big)~,\\
W_3^{\text{loc}}&=\frac{1}{6}\Big( \vev{\Tr \Phi}\big|_{\text{loc}}^3- 
3 \vev{\Tr \Phi}\big|_{\text{loc}}\vev{\Tr \Phi^2}\big|_{\text{loc}} + 2 \vev{\Tr \Phi^3}\big|_{\text{loc}}\Big)~,\\
W_4^{\text{loc}}&=\frac{1}{24}\Big(\vev{\Tr \Phi}\big|_{\text{loc}}^4 - 
6 \vev{\Tr \Phi}\big|_{\text{loc}}^2 \vev{\Tr \Phi^2}\big|_{\text{loc}} + 3 \vev{\Tr \Phi^2}\big|_{\text{loc}}^2 \\
&\qquad\qquad+ 8 \vev{\Tr \Phi}\big|_{\text{loc}} \vev{\Tr \Phi^3}\big|_{\text{loc}} - 
6 \vev{\Tr \Phi^4}\big|_{\text{loc}}\Big)~,
\end{aligned}
\end{equation}
and so on. Plugging the localization results (\ref{Trphin}), after some long but straightforward algebra, we find
\begin{align}
W_1^{\text{loc}}&=\sum_{u}a_u~,\label{W1loc}\\
W_2^{\text{loc}}&=\sum_{u_1<u_2}a_{u_1}a_{u_2} -\frac{N\,m^2}{24}\big(E_2-1\big)
+\frac{m^4}{288}\big(E_2^2-E_4\big)\,C^{\,0}_{\,2}\label{W2loc}\\
&~~+\frac{m^6}{4320}\big(5E_2^3-3E_2E_4-2E_6\big)\,\,C^{\,0}_{\,4}
+\frac{m^6}{3456}\big(E_2^3-3E_2E_4+2E_6\big)\,\,C^{\,0}_{\,2;11}+O(m^8)~,\phantom{\Bigg|}\notag\\
W_3^{\text{loc}}&=\!\!\!\!\sum_{u_1<u_2<u_3}\!\!\!\!a_{u_1}a_{u_2}a_{u_3}-
\frac{(N-2)\,m^2}{24}\,\big(E_2-1\big)\sum_{u}a_u+\frac{m^4}{288}\big(E_2^2-E_4\big)
\Big(C^{\,0}_{\,2}\sum_{u}a_u-2C^{\,1}_{\,2}\Big)\notag\\
&~~+\frac{m^6}{4320}\,\big(5E_2^3-3E_2E_4-2E_6\big)\Big(C^{\,0}_{\,4} \,\sum_{u} a_u-2\, C^{\,1}_{\,4}\Big)\notag\\
&~~+\frac{m^6}{3456}\,\big(E_2^3-3E_2E_4+2E_6\big)\Big(C^{\,0}_{\,2;11} \,
\sum_{u} a_u-2\, C^{\,1}_{\,2;11}\Big)+O(m^8)~,\phantom{\Bigg|}\label{W3loc}\\
W_4^{\text{loc}}&=\!\!\!\sum_{u_1<\cdots<u_4}\!\!a_{u_1}a_{u_2}a_{u_3}a_{u_4}
-\frac{m^2}{24}\,\big(E_2-1\big)\Big(\big(\sum_{u}a_u\big)^2+(N-6)\sum_{u_1<u_2}a_{u_1}a_{u_2} \Big)\notag\\
&~~+\frac{m^4}{288}\,(E_2^2-E_4)\Big(C^{\,0}_{\,2} \sum_{u_1<u_2} a_{u_1} a_{u_2} -
2\,C^{\,1}_{\,2} \,\sum_u a_u + 3\,C^{\,2}_{\,2}-\binom{N}{2}\Big)\notag\\
&~~+\frac{N\,m^4}{5760}\Big(5N\big(3E_2^2-2E_4-2E_2+1\big) - 30 E_2^2+ 12 E_4+60 E_2 -42\Big)\notag\\
&~~+\frac{m^6}{4320}\,\big(5E_2^3-3E_2E_4-2E_6\big)
\Big(C^{\,0}_{\,4}\sum_{u_1<u_2} a_{u_1} a_{u_2} -2\, C^{\,1}_{\,4}\,
\sum_{u} a_u+3\,C^{\,2}_{\,4}-\frac{1}{2}\,C^{\,0}_{\,2}\Big)\notag\\
&~~+\frac{m^6}{3456}\,\big(E_2^3-3E_2E_4+2E_6\big)
\Big(C^{\,0}_{\,2;11}\sum_{u_1<u_2} a_{u_1} a_{u_2} -2\, C^{\,1}_{\,2;11}\,\sum_{u} a_u+3\,C^{\,2}_{\,2;11} 
\Big)\notag\\
&~~-\frac{(N-6)\,m^6}{6912}\,\big(E_2-1\big)\big(E_2^2 - E_4\big)
\,C^{\,0}_{2}+O(m^8)~.\phantom{\Bigg|}\label{W4loc}
\end{align}
It is remarkable to see in these expressions the same combinations of Eisenstein series and of lattice sums 
appearing in the $W_n$ presented in (\ref{W1q})--(\ref{W4q}). 
However, there are also some important differences which we are going to discuss.

The first observation is that, even though the classical part of the $W_n^{\text{loc}}$ 
is the degree $n$ symmetric polynomial in the vacuum expectation values, the full 
$W_n^{\text{loc}}$ do not satisfy the 
corresponding chiral ring relations.\footnote{This was already noted in \cite{Cachazo:2002ry}\cite{Flume:2004rp}\cite{Nekrasov:2012xe} for pure $\mathcal{N}=2$ SYM theories.} Indeed, it is not difficult to verify that
\footnote{Recall that the localization formulas formally hold true also for $N=1$.}
\begin{equation}
W_{2,3,4}^{\text{loc}}\Big|_{\mathrm{U}(1)}\neq 0~,\qquad
W_{4}^{\text{loc}}\Big|_{\mathrm{U}(2)}\neq 0~,\qquad
W_{4}^{\text{loc}}\Big|_{\mathrm{U}(3)}\neq 0~,
\label{ineq}
\end{equation}
whereas in all these cases one should expect a vanishing result if the $W_{n}^{\text{loc}}$ were
the quantum version of the classical symmetric polynomials.
We find that enforcing the chiral ring relations allows us to make contact with the results for the $W_n$ coming from the Seiberg-Witten curves. This amounts a redefinition of $W_n^{\text{loc}}$, and thereby a different choice of the generators for the chiral ring. 

The second observation is that our explicit localization results allow us to perform this redefinition
in a systematic way. Indeed, {from}
\begin{equation}
W_{2}^{\text{loc}}\Big|_{\mathrm{U}(1)}=-\frac{m^2}{24}\big(E_2-1\big)~,
\end{equation}
we immediately realize that the ``good'' operator at level 2 can be obtained from 
$W_{2}^{\text{loc}}$ by removing the constant $m^2$ term proportional to $(E_2-1)$. 
We are thus led to define\,%
\footnote{It is interesting to note that also the prepotential of $\mathcal{N}=2^\star$ theories
satisfies the duality properties discussed in \cite{Billo':2015ria,Billo':2015jta} only 
if an $a$-independent term proportional to $m^2$, which is not quasi-modular, is discarded. 
Such a constant term in the prepotential does not, however, influence the effective action.}
\begin{equation}
\widehat{W}_2= W_2^{\text{loc}}+ \frac{N\,m^2}{24}\,\big(E_2-1)~.
\label{What2}
\end{equation}
Similarly, at level 3 we find that the term responsible for the inequalities in (\ref{ineq}) is again the $m^2$
part proportional to $(E_2-1)$, so that the desired operator is
\begin{equation}
\widehat{W}_3= W_3^{\text{loc}}+ \frac{(N-2)\,m^2}{24}\,\big(E_2-1)\sum_{u}a_u~.
\label{What3}
\end{equation}
At level 4 we see that the non-vanishing results in (\ref{ineq}) are due again to the $m^2$ terms
proportional to $(E_2-1)$ but also to the $a$-independent terms at order $m^4$ and to the $m^6$ terms
in the last line of (\ref{W4loc}). This motivates us to introduce
\begin{equation}
\begin{aligned}
\widehat{W}_4&= W_4^{\text{loc}}+\frac{m^2}{24}\,\big(E_2-1\big)\Big(\big(\sum_{u}a_u\big)^2
+(N-6)\sum_{u_1<u_2}a_{u_1}a_{u_2} \Big)\\
&~~-\frac{N\,m^4}{5760}\Big(5N\big(3E_2^2-2E_4-2E_2+1\big) - 30 E_2^2+ 12 E_4+60 E_2 -42\Big)\\
&~~+\frac{(N-6)\,m^6}{6912}\,\big(E_2-1\big)\big(E_2^2 - E_4\big)
\,C^{\,0}_{2}~.\phantom{\Bigg|}
\end{aligned}
\label{What4}
\end{equation}
It is interesting to observe that the difference between $\widehat{W}_n$ and $W_n^{\text{loc}}$ only
consists of  terms whose coefficients are polynomials in the Eisenstein series that do not have a definite
modular weight, whereas the common terms at order $m^{2\ell}$ are quasi-modular forms 
of weight $2\ell$. Removing all such inhomogeneous terms from the $W_n^{\text{loc}}$ yields 
the one-point functions that satisfy the classical chiral ring relations. Furthermore, it is worth noticing that 
(\ref{What4}) can be rewritten as
\begin{equation}
\begin{aligned}
\widehat{W}_4&=W_4^{\text{loc}}+\frac{(N-6)\,m^2}{24}\,\big(E_2-1\big) W_2^{\text{loc}}
+\frac{m^2}{24}\,\big(E_2-1\big) W_1^2\\
&~~-\frac{N\,m^4}{5760}\,\Big(5N\big(E_2^2-2E_4+2E_2-1\big) +30 E_2^2+ 12 E_4-60 E_2 +18\Big)~.\label{diff4}
\end{aligned}
\end{equation}
The fact that the $m^6$ terms are exactly reabsorbed is a very strong indication that 
the above formula is exact in $m$. Notice also that this redefinition, like 
the previous ones (\ref{What2}) and
(\ref{What3}), is exact in the gauge coupling.

The most important point, however, is that the resulting expressions for the $\widehat{W}_n$ derived
from the localization formulas precisely match those for the $W_n$ obtained from 
the SW curves in the previous section. Indeed,
comparing (\ref{What2})--(\ref{What4}) with (\ref{W2q})--(\ref{W4q}), we have
\begin{equation}
\widehat{W}_n=W_n~.
\label{equality}
\end{equation}
Our calculations provide an explicit proof of this equivalence for $n\leq 4$, but of course
they can be generalized to higher levels. 

Summarizing, we have found that the quantum coordinates of the moduli space 
computed using the SW curves for the $\mathcal{N}=2^\star$ U($N$) theory
agree with those obtained from the localization formulas provided on the latter we enforce the classical
chiral ring relations obeyed by the symmetric polynomials. 
Enforcing these relations is clearly a choice that amounts to selecting a particular
basis for the generators of the chiral ring.
It would be interesting to explore the possibility of modifying the localization
prescription in order to obtain chiral observables that automatically satisfy such relations without the
need for subtracting the non-quasi-modular terms. 

\section{1-instanton results}
\label{secn:one-inst}
In the previous sections we have presented a set of results that are exact in the gauge coupling 
constant for quantities that have been evaluated order by order in the hypermultiplet mass. 
Here instead, we exhibit a result that is exact in $m$ but is valid only at the 1-instanton level. 
To do so let us consider the localization results (\ref{Trphinq}) for the one-point functions 
$\vev{\Tr \Phi^n}\big|_{\text{loc}}$,
and focus on the terms proportional to $q$ corresponding to $k=1$. Actually, the calculations at $k=1$ can be
easily performed also for higher rank groups and pushed to higher order in the mass without any problems.
Collecting these results, it is does not take long to realize that
they have a very regular pattern and can be written compactly as
\begin{equation}
\begin{aligned}
\vev{\Tr \Phi^n}\big|_{k=1}&=-n(n-1)\,q\,m^2\bigg(C^{\,n-2}-m^2\,\,C^{\,n-2}_{\,2}-\frac{m^4}{2}
\,C^{\,n-2}_{\,2;11}-\frac{m^6}{24}\,C^{\,n-2}_{\,2;1111}+\cdots\bigg)\\
&=-n(n-1)\,q\,m^2\bigg(C^{\,n-2}-\sum_{\ell=0}\frac{m^{2+\ell}}{\ell!}\,\,
C^{\,n-2}_{2;\underbrace{\mbox{\scriptsize{1\ldots1}}}_{\mbox{\scriptsize{$\ell$}}}}\bigg)~.
\end{aligned}
\label{1inst}
\end{equation}
Notice that $C^{\,p}_{\,2;1\cdots1}$ with an odd number of 1's is zero, and that for a U($N$) theory
only $N-1$ terms are present in the sum over $\ell$. Using the explicit form of the
lattice sums (\ref{Cpnm}), one can resum the above expression and find
\begin{equation}
\vev{\Tr \Phi^n}\big|_{k=1}=-n(n-1)\,q
\,m^2\sum_{\lambda\in\mathcal{W}}(\lambda\cdot \phi)^{n-2}\bigg[1-
\sum_{\alpha\in\Psi_\lambda}\frac{m^2}{(\alpha\cdot\phi)^2}\prod_{\beta\in\Psi_\alpha}\Big(1+\frac{m}{\beta\cdot\phi}\Big)\bigg]~.
\label{1instres}
\end{equation}
This is a generalization of an analogous formula for the prepotential found in 
\cite{Billo':2015ria,Billo':2015jta}, 
to the case of the chiral observables of the $\mathcal{N}=2^\star$ theory. 
Being exact in $m$, we can use (\ref{1instres}) to decouple the hypermultiplet by sending its mass
to infinity and thus obtain the
1-instanton contribution to the one-point function of the single trace operators in 
the pure $\mathcal{N}=2$ U($N$) gauge theory.
More precisely, this decoupling limit is 
\begin{equation}
m\to\infty~~\mbox{and}~~q\to 0 \quad\mbox{with}~~q\,m^{2N}\equiv \Lambda^{2N}~~\mbox{fixed}~.
\label{decoupling}
\end{equation}
Recalling that the number of roots $\beta$ in $\Psi_{\alpha}$ is $2N-4$, we see that the highest mass 
power in (\ref{1instres}) is precisely $m^{2N}$, so that in the decoupling limit we get
\begin{equation}
\vev{\Tr \Phi^n}\big|_{k=1}=\,n(n-1)\,\Lambda^{2N}\sum_{\lambda\in\mathcal{W}}
\sum_{\alpha\in\Psi_\lambda}\frac{(\lambda\cdot \phi)^{n-2}}{(\alpha\cdot\phi)^2}
\prod_{\beta\in\Psi_\alpha}\frac{1}{\beta\cdot\phi}~.
\label{1instpure}
\end{equation}
We remark that for $n=2$ this formula agrees with the 1-instanton prepotential of the 
pure $\mathcal{N}=2$ theory, which was derived in \cite{Benvenuti:2010pq,Keller:2011ek} 
using completely different methods. Indeed, 
through the Matone relation \cite{Matone:1995rx} $\vev{\Tr \Phi^2}$ and the prepotential 
at 1 instanton 
are proportional to each other. 

Moreover, if we restrict to SU($N$), it is possible to verify that
(\ref{1instpure}) is in full agreement with the chiral ring relations of the pure $\mathcal{N}=2$
SYM theory that follow by expanding in inverse powers of $z$ the identity \cite{Cachazo:2002ry} \cite{Flume:2004rp} \cite{Nekrasov:2012xe}
\begin{equation}
\Big\langle\mathrm{Tr}\,\frac{1}{z-\Phi}\Big\rangle=\frac{P_N^\prime(z)}{\sqrt{P_N^2(z)-
4\Lambda^{2N}}}
\end{equation}
where 
\begin{equation}
P_N(z)=z^N+\sum_{\ell=2}^Nu_\ell\,z^{N-\ell}\,,
\end{equation}
is a degree $N$ polynomial that encodes the Coulomb moduli $u_\ell$ appearing in the SW curve of the pure SU($N$) SYM theory.

It would be nice to see whether the formulas (\ref{1instres}) and (\ref{1instpure}) for generic $n$ 
are valid also for other groups, as is the case for the $n=2$ case \cite{Keller:2011ek}\cite{Billo':2015ria,Billo':2015jta}.

\section{Conclusions and discussion}
\label{concl}

In this work we have performed a detailed analysis of the simplest chiral observables 
constructed from the adjoint scalar $\Phi$ of
the ${\mathcal N}=2^{\star}$ U($N$) SYM theory.
The expressions for $\vev{\Tr \Phi^n}$ that we obtained using localization methods are written as 
mass expansions, with the dependence on the gauge coupling constant being completely resummed 
into quasi-modular forms, and the dependence on the classical vacuum expectation values 
expressed through lattice sums involving the roots and weights of the gauge algebra.  
Therefore, these findings can be thought of as a natural generalization of the results 
obtained in \cite{Billo':2015ria}--\nocite{Billo':2015jta}\cite{Billo:2016zbf} for the prepotential 
to other observables of the ${\mathcal N}=2^{\star}$ theory.

We also found that the symmetric polynomials $W_n$ constructed out of $\vev{\Tr \Phi^n}$ do not 
satisfy the classical chiral ring relations \cite{Nekrasov:2012xe}, while some simple redefinitions allow one 
to enforce them. The redefined chiral observables obtained in this way perfectly match those we derived 
by completely independent means, namely from the SW curves and 
the associated period integrals, or from modular anomaly equations. 
We then identified particular combinations $A_n$ of chiral observables that transform as modular 
forms of weight $n$ under the non-perturbative S-duality group, and derived a 
relation between the $W_n$ and the $A_n$ which is exact both in the hypermultiplet mass
and in the gauge coupling constant. 

Given that our results are a generalization of what was found 
in \cite{Billo':2015ria}--\nocite{Billo':2015jta}\cite{Billo:2016zbf}, it is natural to ask ourselves
about the possibility of extending the above analysis to ${\mathcal N}=2^{\star}$ theories with 
other classical groups. In this respect we recall that the integrable system that governs 
the quantum gauge theory for these cases and the associated Lax pair have been 
obtained in \cite{D'Hoker:1998yg,D'Hoker:1998yi}. 
However, for the $D_n$ series, the explicit form of the spectral curves in terms of elliptic and modular forms 
is only known for cases with low rank \cite{D'Hoker:1999ft}. 
Thus, it would be very interesting to revisit this problem in the present context, especially given 
the significant progress that has been made relating gauge theories and integrable systems 
over the past decade \cite{Nekrasov:2009uh, Nekrasov:2009ui, Nekrasov:2009rc}. 
The localization results available for a generic group $G$ would provide additional checks 
on the correctness of the proposed solution. Another important class of theories to consider would 
be the superconformal ADE quiver-type models studied in 
\cite{Nekrasov:2012xe} and their $\Omega$-deformed generalizations \cite{Nekrasov:2013xda}.  

It would also be worthwhile to calculate these chiral observables for other theories, such as SQCD-like theories. In these cases, the prepotential has been resummed in terms of quasi-modular forms of generalized triangle groups in a special locus on the 
moduli space \cite{Ashok:2015cba, Ashok:2016oyh} and thus it would be interesting to see if one can obtain 
similar results for the one point functions of chiral observables as well.

Finally, we remark that the calculation of the one point functions $\vev{\Tr \Phi^n}$
has an important role in the physics of surface operators \cite{Gukov:2006jk,Gukov:2008sn} (for
a review see for instance \cite{Gukov:2014gja}). The infrared physics of surface operators 
in ${\mathcal N}=2$ gauge theories is in fact captured 
by a twisted effective superpotential in a two dimensional theory. As shown in \cite{Gaiotto:2013sma},
one of the ways in which this twisted superpotential can be determined is from the generating function 
of the expectation values of chiral ring elements in the bulk four dimensional theory. Our results can
be interpreted as a first step in this direction. 
Furthermore, it would be interesting to explore if the existence of 
combinations of chiral ring elements that have simple modular behaviour under S-duality can be useful 
to improve our understanding of the two dimensional theory that captures the infrared physics of surface operators.

\vskip 1.5cm
\noindent {\large {\bf Acknowledgments}}
\vskip 0.2cm
We thank F.~Fucito, L.~Gallot, D.~Jatkar, R.~R.~John, R.~Loganayagam, J.~F.~Morales, J.~Troost, and A.~Zein Assi for discussions. S.A.~and E.D.~would like to thank the University of Turin and INFN, Turin for hospitality during the completion of this work.

The work of M.B., M.F., and M.M.~is partially supported by the Compagnia di San Paolo 
contract ``MAST: Modern Applications of String Theory'' TO-Call3-2012-0088.
\vskip 1 cm

% Begin Back Matter -----------------------------------------------------------
\appendix

\section{Eisenstein series and elliptic functions}
\label{appeisen}
\subsection*{$\bullet$ Eisenstein series}
The Eisenstein series $E_{2n}$ are holomorphic functions of $\tau\in \mathbb{H}_+$ defined as
\begin{equation}
\label{defeis}
E_{2n} = \frac{1}{2\zeta(2n)}\sum_{m,n\in\mathbb{Z}^2\setminus \{0,0\}} \frac{1}{(m+n\tau)^{2n}}~.
\end{equation}
For $n>1$, they are modular forms of weight $2n$, namely under an $\mathrm{SL}(2,\mathbb{Z})$ transformation
\begin{equation}
\label{sl2z}
\tau \to \tau^\prime = \frac{a\tau + b}{c\tau +d}~~~\mbox{with}~~
a,b,c,d \in\mathbb{Z}~~~\mbox{and}~~
ad-bc=1~,
\end{equation}
they transform as
\begin{equation}
\label{eissl2z}
E_{2n}(\tau^\prime) = (c\tau + d)^{2n} E_{2n}(\tau)~.
\end{equation}
For $n=1$, the $E_2$ series is instead quasi-modular. Its modular transformation has in fact an anomalous term:
\begin{equation}
\label{eis2sl2z}
E_{2}(\tau^\prime) = (c\tau + d)^{2} E_{2}(\tau) + \frac{6}{\ii\pi}c(c\tau + d)~.
\end{equation}
All modular forms of weight $2n>6$ can be expressed as polynomials of $E_4$ and $E_6$; 
the quasi-modular forms instead can be expressed as polynomials in $E_2$, $E_4$ and $E_6$.

The Eisenstein series admit a Fourier expansion in terms of $q={\mathrm{e}}^{2\pi\ii\tau}$ of the form 
\begin{equation}
E_{2n}=1+\frac{2}{\zeta(1-2n)}\sum_{k=1}^\infty \sigma_{2n-1}(k) q^{k}~,
\label{eis2nq0}
\end{equation}
where $\sigma_p(k)$ is the sum of the $p$-th powers of the divisors of $k$. 
In particular, this amounts to
\begin{equation}
\begin{aligned}
E_2&=1-24\sum_{k=1}^\infty\sigma_1(k) q^k=1-24q-72q^2-96q^3+\cdots~,\\
E_4&=1+240\sum_{k=1}^\infty\sigma_3(k) q^k=1+240q+2160q^2+6720q^3+\cdots~,\\
E_6&=1-504\sum_{k=1}^\infty\sigma_5(k) q^k=1-504q-16632q^2-122976q^3+\cdots~.
\end{aligned}
\end{equation}
The quasi-modular and modular forms are connected to each other by logarithmic $q$-derivatives as
\begin{equation}
q \frac{d E_2 }{dq} = \frac{1}{12} \left(E_2^2 - E_4\right)~,\quad
q \frac{d E_4}{dq}   = \frac{1}{3}  \left(E_2 E_4 - E_6\right)~,\quad
q \frac{d E_6}{dq}  = \frac{1}{2}  \left(E_2 E_6 - E_4^2\right)~,
\label{De2e4d6}
\end{equation}
while $E_2$ is related to the derivative of the Dedekind $\eta$-function
\begin{equation}
\label{etadef}
\eta(q) = q^{1/24}\prod_{k=1}^\infty (1 - q^{k})~.
\end{equation}
In fact, we have
\begin{equation}
\label{Dlogeta}
q \frac{d}{dq} \log\left(\frac{\eta}{q^{1/24}}\right) = - \sum_{k=1}^\infty \sigma_1(k) q^{k} =
\frac{E_2 - 1}{24}~. 
\end{equation}
\subsection*{$\bullet$ $\theta$-functions}
The Jacobi $\theta$-functions are defined as 
\begin{equation}
 \theta\left[^a_b\right](z\vert \tau) = \sum_n \mathrm{e}^{\pi\ii\tau \left(n - \frac{a}{2}\right)^2
+ 2\pi\ii\left(z-\frac{b}{2}\right)\left(n - \frac{a}{2}\right)}~,
\label{vart}
\end{equation}
for $a,b=0,1$. 
These functions are quasi-periodic, in a multiplicative fashion, for shifts of the variable  
$z$ by a lattice element $\lambda = p \tau + q$, with $p, q \in \mathbb R$; in fact one has
\begin{equation}
\label{qpth}
\theta\left[^a_b\right](z+\lambda\vert  \tau) = e(\lambda,z) \, \theta\left[^a_b\right](z\vert \tau)~,
\end{equation}
where 
\begin{equation}
e(\lambda,z) = \mathrm{e}^{-\pi\ii\tau p^2 - 2 \pi\ii p \left(z-\frac{b}{2}\right)-
 \pi\ii \,a\,q}~.
\end{equation}
As customary, we use the notation
\begin{equation}
\begin{aligned}
\theta_1(z\vert \tau)&=\theta\left[^1_1\right](z\vert \tau)~,~~~~
\theta_2(z\vert \tau)=\theta\left[^1_0\right](z\vert \tau)~,\\
\theta_3(z\vert \tau)&=\theta\left[^0_0\right](z\vert \tau)~,~~~~
\theta_4(z\vert \tau)=\theta\left[^0_1\right](z\vert \tau)~.
\end{aligned}
\label{t1234} 
\end{equation}
By evaluating these functions at $z=0$, one obtains the so-called 
$\theta$-constants $\theta_a(\tau)$,
which satisfy the abstruse identity:
\begin{equation}
 \theta_3(\tau)^4-\theta_2(\tau)^4-\theta_4(\tau)^4=0~,
\label{idas}
\end{equation}
while $\theta_1(\tau)=0$.

The Eisenstein series $E_4$ and $E_6$ can be written as polynomials in the $\theta$-constants according to
\begin{equation}
\begin{aligned}
E_4 &= \frac{1}{2} \big(\theta_2(\tau)^8 + \theta_3(\tau)^8+ \theta_4(\tau)^8\big) ~,\\
E_6 &=  \frac{1}{2} \big(\theta_3(\tau)^4 + \theta_4(\tau)^4\big)\big(\theta_2(\tau)^4 
+ \theta_3(\tau)^4\big)\big(\theta_4(\tau)^4 - \theta_4(\tau)^4\big)~.
\end{aligned}
\label{e4e6theta}
\end{equation}

\subsection*{$\bullet$ Weierstra\ss~function}
The Weierstra\ss~function $\wp(z\vert \tau)$ defined by
\begin{equation}
\label{defwpZ}
\wp(z\vert \tau) = \frac{1}{z^2} + \sum_{m,n\in\mathbb{Z}^2\setminus \{0,0\}} \left(\frac{1}{(z + m\tau   + n)^2} -
\frac{1}{( m\tau + n)^2}\right)~,
\end{equation}
is a meromorphic function in the complex $z$-plane with a double pole in $z=0$, which is 
doubly periodic with periods $1$ and $\tau$. We often leave the $\tau$-dependence implicit, 
and write simply $\wp(z)$. 

It is  a Jacobi form of weight 2 and index 0, namely under a modular transformation~\eqref{sl2z} 
combined with $z\to z'=z/(c\tau+d)$, it transforms as
\begin{equation}
  \wp(z'\vert\tau')=(c\tau+d)^2\wp(z\vert \tau) \, .
  \label{wpmodular}
\end{equation}
It also satisfies the following differential equation
\begin{equation}
\wp'(z\vert \tau)^2 = 4 \,\wp^3(z\vert \tau) -\frac{4\pi^4\,E_4}{3}\, \wp(z\vert \tau)
 -\frac{8\pi^6\,E_6}{27}~.
\label{PWapp}
\end{equation}

Using the quasi-periodicity properties of the $\theta$-functions given in (\ref{qpth}),
it is easy to show that second derivative of $\theta_1$ is a proper periodic function; indeed
\begin{equation}
\label{d2logper}
\frac{d^2}{dz^2}\log\theta_1(z+m+n\tau\vert \tau) = \frac{d^2}{dz^2}\log\theta_1(z\vert \tau)~.
\end{equation}
Furthermore, by studying its pole structure, it is possible to show that it coincides 
with the Weierstra\ss~function, up to a $z$-independent term:
\begin{equation}
\label{Pisd2logt1}
\wp(z\vert \tau) = - \frac{d^2}{dz^2}\log\theta_1(z|\tau)+ c~.
\end{equation} 
The explicit evaluation of the constant shows that
\begin{equation}
\label{cres}
c  = -\frac{\pi^2}{3} \left(1 - 24 \sum_{k=1}^\infty \frac{q^k}{(1 - q^k)^2}\right)~=
-\frac{\pi^2}{3}\left(1-24\,\sum_{k=1}^\infty \sigma_1(k) \,  q^k\right)~=-\frac{\pi^2}{3} E_2~,
\end{equation}
so that we have
\begin{equation}
\label{wpise2}
\wp(z\vert \tau) =  - \frac{d^2}{dz^2}\log\theta_1(z|\tau)  -\frac{\pi^2}{3} E_2~.
\end{equation}
Using the notation of Section~\ref{secn:DPcurve} (see in particular (\ref{hn})), 
from (\ref{wpise2}) one can easily show that
\begin{equation}
h_1^\prime=\frac{1}{2\pi\ii}\,\frac{d}{dz}h_1(z)=
\frac{1}{(2\pi\ii)^2}\frac{d^2}{dz^2}\log\theta_1(z|\tau)=-\frac{\wp(z\vert\tau)}{(2\pi\ii)^2}+\frac{E_2}{12}
\label{proofx}
\end{equation}
which proves the first identity in (\ref{rel}). By taking further derivatives of this equation
with respect to $2\pi\ii z$ and using the differential equation (\ref{PWapp}), 
one can straightforwardly prove the other identities in
(\ref{rel}).

Using the periodicity property (\ref{d2logper}), it is possible to exploit the relation (\ref{wpise2}) to
deduce the values of the integral of the $\wp$ function 
along the $\alpha$ and $\beta$ cycles of the torus, that are parametrized respectively by 
$z=\gamma$ and $z=\gamma \tau$, with $\gamma \in [0,1]$; for instance we have
\begin{equation}
\oint_{\alpha} \wp(z\vert \tau)=  
-\frac{\pi^2}{3}\,E_2~.
\label{intP1}
\end{equation}
This result has been used in Section~\ref{secn:periodsandmodularity}, see in particular (\ref{intalpha}).

By differentiating  the differential equation (\ref{PWapp}) and using the previous result, one
can compute also the integral of higher powers of $\wp$. For instance, the first derivative of
(\ref{PWapp}) yields the relation
\begin{equation}
\wp(z\vert \tau)^{\prime\prime}=6 \,\wp(z\vert \tau)^2-\frac{2\pi^4}{3}\,E_4
\end{equation}
{from} which we find
\begin{equation}
\oint_{\alpha} \wp^2(z\vert \tau)= \frac{\pi^4}{9}\,E_4~.
\label{intP2}
\end{equation}
Proceeding in this way, one can easily compute the period integrals for higher powers of $\wp$, 
(see for example \cite{KashaniPoor:2012wb} and references therein).

\section{Generalized Donagi-Witten polynomials}
\label{DWpolygeneral}

In Section~\ref{DW} we obtained the expression of the first polynomials $P_n$ that appear in the 
Donagi-Witten curve, by imposing the requirements that they satisfy the recursion relation 
\begin{equation}
\frac{dP_n}{dt} = n P_{n-1} ~,
\label{rP}
\end{equation}
and that their behaviour at infinity is
\begin{equation}
P_n\Big(t+\frac{m}{u}\Big) \sim \frac{\alpha_n}{u} +\mathrm{regular}~.
\end{equation}
This procedure can be iteratively carried out order by order in $n$. 
The general form of the $P_n$ required from (\ref{rP}) is
\begin{equation}
P_n = t^n - \sum_{p=2}^n (-1)^p\, (p-1)\, x_p\, m^{p} \binom{n}{p} t^{n-p} ~,
\label{generalPn}
\end{equation}
where the coefficients $x_p$ are elliptic and modular forms of weight $p$ that can be fixed recursively.
As discussed in the main text, up to $n=3$ the solution to the constraints 
is unique, namely
\begin{equation}
\begin{aligned}
P_0 &=1~,\qquad P_2 = t^2 - m^2\,x~,\\
P_1 &=t~,\qquad \,P_3 = t^3 - 3\,t\, m^2\,x +2 m^3\,y ~.
\end{aligned}
\label{p123}
\end{equation}
{From} $n=4$ on, several combinations of elliptic and modular forms start to appear and their 
relative coefficients are not uniquely fixed by the requirement of the behaviour at infinity. 
For instance, for $n=4$ and $n=5$ one finds a one-parameter family of solutions, and
for $n=6$ a two-parameter family of solutions, given by
\begin{equation}
\begin{aligned}
P_4 &=t^4-6\,t^2\,x\,m^2+8\,t\,y\,m^3-\left(3\,x^2-\alpha\,E_4\right)m^4~,\\
P_5&=t^5 -10\,t^3\,m^2 x + 20\, t^2\,m^3 y-5\,t\left(3\,x^2-\alpha\,E_4\right)m^4+4 \,m^5\, x\,y~,\\
P_6&= t^6-15\,t^4\, m^2x + 40\, t^3\, m^3 y -15\,t^2\left(3\,x^2-\alpha\,E_4\right)m^4+24\,t\, 
m^5\, x\,y \\
&\qquad-\,m^6
\Big(\big(5+\beta\big)\,x^3- \beta\,y^2- \frac{E_4}{48}\,\big(32-720\,\alpha +\beta\big)\,x\Big)  ~.
\end{aligned}
\label{p456}
\end{equation}
These polynomials correspond to the expression in (\ref{generalPn}) where the first few $x_p$ are
\begin{equation}
\begin{aligned}
x_2 &= x ~,\qquad x_3 = y ~,\qquad x_4 = x^2 -\frac{\alpha}{3} E_4~,\\ 
x_5 &= xy~, \qquad x_6 = \frac{1}{5}\Big(\big(5+\beta\big)\,x^3- \beta\,y^2- \frac{E_4}{48}\,\big(32-720\,\alpha +\beta\big)\,x\Big) ~.
\end{aligned}
\label{explicitxk}
\end{equation}

\section{Modular covariance from the D'Hoker-Phong curve}
\label{secn:modularDP}

In this Appendix we explain how to obtain the relation \eqref{eq:Akresult} between
the modular covariant $A_n$ and the $W_n$, directly from the D'Hoker-Phong form of the SW 
curve instead of comparing it with the Donagi-Witten curve as we did in Section~\ref{secn:curvecomp}.

Recall that in the D'Hoker-Phong approach the SW curve is given by
\begin{align}
R(t,z)=&\sum_{\ell=0}^N (-1)^\ell\,W_\ell \left[ t - 
m \left( \frac{1}{2\pi\ii}\frac{d}{d z} + h_1(z) \right) \right]^{N-\ell} 1~\Bigg\vert_{h_1=0}=0
  \label{eq:Rrewriteapp}
\end{align}
As discussed in the main text, the coefficients $W_\ell$ do not transform 
homogeneously under S-duality. One can see this clearly by analyzing how the 
other objects appearing in \eqref{eq:Rrewriteapp} transform.
In fact, using \eqref{proofx}, the modular property~\eqref{wpmodular} of the Weierstra\ss~function 
implies that $h_1'$ transforms as a quasi-modular form of weight 2.
Acting with additional derivatives on both sides of \eqref{proofx} kills the term proportional to $E_2$
so that the $n$-th derivative of $h_1$ for $n>1$ transforms homogeneously with weight $n+1$.
On the other hand, from the analysis in section~\ref{DW}, we know that one can rewrite the equation 
for the curve such that it becomes modular of weight $N$.
Hence there must exist some inhomogeneous transformation law of the $W_n$, compensating the 
inhomogeneous transformation of $h_1'$, such that the whole polynomial is modular covariant.
Indeed, if not for this inhomogeneous transformation of $h_1'$, the curve would be manifestly modular covariant.

These observations suggest to introduce a new function $R_{\text{mod}}(t,z)$ with coefficients 
$A_\ell$, by substituting the quasi-modular $h_1'$ for the modular expression $h_1'-E_2/12$, namely
\begin{equation}
R_{\text{mod}}(t,z)=\sum_{\ell=0}^N (-1)^\ell\,A_\ell \left[ t - 
m \left( \frac{1}{2\pi\ii}\frac{d}{d z} + h_1(z) \right) \right]^{N-\ell} 
1~\Bigg\vert_{h_1=0\,,\,h_1'\to h_1'-E_2/12}~.
\label{eq:Rmoddef}
\end{equation}
By construction, this polynomial is modular of weight $N$ if the coefficients $A_\ell$ are modular 
of weight $\ell$. Equating $R_{\text{mod}}=R$ then yields a relation between the modular covariant 
$A_\ell$ and the expectation values of symmetric polynomials $W_\ell$, 
which agrees exactly with \eqref{eq:Akresult}. In fact, the asymptotic 
expansion at large $t$ of $R_{\text{mod}}$ reads
\begin{equation}
\begin{aligned}
R_{\text{mod}}(t,z) =& \,t^N -t^{N-1}\, A_1
+ t^{N-2}\left[A_2 +\binom{N}{2}\,m^2\left(h_1^\prime-\frac{E_2}{12}\right) \right] \\
&-t^{N-3}\left[A_3 +\binom{N-1}{2}\,m^2\left(h_1'-\frac{E_2}{12}\right)A_1
+m^3\binom{N}{3}h_1''\right]\\
&+t^{N-4}\left[A_4 +\binom{N-2}{2}\,m^2\left(h_1'-\frac{E_2}{12}\right) A_2
+m^3\binom{N-1}{3}\,h_1''\,A_1\right.\\
&~~~~~\quad\quad\left.+\binom{N}{4}\,m^4\left(h_1^{(3)}+3
\left(h_1'-\frac{E_2}{12}\right)^2\right) \right] + O(t^{N-5}) ~.
\end{aligned}
\end{equation}
By comparing this with (\ref{eq:Rrewrite}) and equating the coefficients of the various $t$ powers we
can easily find the relation \eqref{eq:Akresult}.

\bibliographystyle{JHEP}

% End Document ----------------------------------------------------------------
\end{document}